\documentclass[draft]{agujournal2019}
\usepackage[utf8]{inputenc}
\usepackage{float}
\usepackage{graphicx}
\usepackage{subcaption}
\usepackage{amsmath}
\usepackage{url} 
\usepackage{lineno}
\usepackage[inline]{trackchanges} 
\usepackage{soul}
\usepackage{natbib}
\draftfalse

\journalname{JGR: Space Physics}

\begin{document}

%
%


\title{How the Shortest and Longest HILDCAAs Shaped Earth’s Outer Radiation Belt During the Van Allen Probes Era?}

%
%




\authors{Ayushi Nema \affil{1},
Ankush Bhaskar\affil{2}, 
Kamlesh N. Pathak \affil{1}, 
Abhirup Datta \affil{3}}

\affiliation{1}{Sardar Vallabhbhai National Institute of Technology, Surat, Gujarat, India, 395007}
\affiliation{2}{Space Physics Laboratory, Vikram Sarabhai Space Centre, ISRO, Trivandrum, India}
\affiliation{3}{Indian Institute of Technology, Indore, Madhya Pradesh, India}




\correspondingauthor{Ayushi Nema}{ayushinema24@gmail.com}



\begin{keypoints}
\item Short-duration HILDCAA events lead to rapid, transient increases in outer radiation belt electron energy and flux, while long-duration events produce sustained and more significant enhancements.
\item A clear correlation was found between increased ULF wave activity and enhanced electron acceleration, whereas VLF waves were also present during the HILDCAAs.
\item These findings emphasize the need to account for the duration of HILDCAA events in space weather models to predict radiation belt dynamics and protect satellite and space-based technologies.
\end{keypoints}

%
%

%
%


\begin{abstract}
High‐intensity long‐duration continuous auroral electrojet (AE) activity (HILDCAA) events are associated with the enhancement of relativistic electron fluxes in the inner magnetosphere. The physical mechanisms underlying this enhancement are not well established yet. In this study, we analyze two contrasting HILDCAA events, one representing the shortest and the other the longest duration, using NASA's Van Allen Probes observations, which have provided unprecedented, unique in-situ observations of the harsh radiation environment around the Earth. Detailed spectral and temporal analyses reveal that while both events trigger enhancements in electron flux across multiple energy channels, the shortest event is characterized by rapid, transient increases in energy levels. In contrast, the longest event produced sudden and markedly higher flux variation. The long duration event showed an acceleration of electrons to higher energy as compared to the shorter one. Moreover, a clear correlation between elevated ULF wave power for the longest event compared to the shortest is observed, apart from chorus waves responsible for relativistic electron acceleration. These findings underscore the importance of considering the duration of events in space weather models and assessment and provide valuable insights into the magnetospheric processes that modulate the variability of the radiation belt during HILDCAA conditions.
\end{abstract}

\section*{Plain Language Summary}
This study investigates how two different space weather events, one very short and one very long, affect Earth’s outer radiation belt. Using data from the Van Allen Probes, we examined changes in electron energy, the number of electrons (flux), and ultra-low frequency (ULF) wave activity during these events. The short event produced a quick, temporary boost in electron energy and flux, while the long event led to a more prolonged and larger increase. We also found that higher ULF wave activity was linked to stronger electron acceleration, suggesting that extended wave-particle interactions during longer events significantly influence the behavior of the radiation belt. These results highlight that the duration of space weather events is a key factor in determining their impact on Earth’s space environment, which is crucial for improving space weather forecasts and protecting satellite and space-based systems.

%
%

\section{Introduction}
High-Intensity Long-Duration Continuous Auroral-Electrojet Activity, also known as HILDCAA events \citep{tsurutani1987cause} are long space weather events in Earth's magnetosphere characterized by continuous auroral activity. It has been known that HILDCAA events are caused during the interaction of high‐speed solar wind streams with slow solar wind streams by forming corotating interaction regions (CIR) \citep{tsurutani2004high, hajra2013solar}. These CIRs, when they interact with the magnetosphere, cause prolonged AE activity. Enhanced Alfvén wave activity in the interplanetary magnetic field (IMF) is also one of the properties of these events. \cite{tsurutani1987cause} have defined HILDCAA events as:[1] High-intensity with peak AE index to be 1000 nT during the period; [2] Long-duration which means events should be atleast 2 days longer; [3] Continuous Auroral Electrojet Activity with AE not dropping below 200 nT for more than 2 hours. HILDCAA events are said to occur outside the main phase of geomagnetic storms \citep{gonzalez1994geomagnetic, russell1974cause}.

Past studies have shown enhanced fluxes of electrons due to HILDCAA activity \citep{hajra2014relativistic,hajra2015relativistic,hajra2024ultra}. This increase in fluxes is primarily attributed to enhanced chorus and ULF wave activities \citep{da2019contribution}. It has been found that the electrons during HIDCAA events are accelerated to relativistic energies via a strong convective electric field at 10 Earth radii in the nightside of the magnetosphere \citep{runov2025themis}. The acceleration of magnetospheric relativistic electrons during HILDCAA events, which are linked to fast solar wind streams coming from coronal holes, is examined in the study by \cite{hajra2024ultra}. One important result from their study is that the maximal energy of the accelerated relativistic electrons increases with the duration of the chorus and HILDCAA events.

PSD represents the number of electrons per unit volume in phase space, providing insights into the processes governing electron acceleration, transport, and loss within the magnetosphere. Accurate PSD analysis is essential for identifying the mechanisms underlying radiation belt electron flux enhancements. To comprehend the acceleration mechanisms of relativistic and ultra-relativistic electrons, PSDs are used as an important quantity \citep{morley2013phase, schiller2017simultaneous}. To  eliminate the adiabatic variations and uncover the non adiabatic electron acceleration, the electron PSDs were expressed as function of the three adiabatic invariants of the particle motion in the geomagnetic field, $\mu$, $K$, and $L^*$ \citep{schulz2012particle}, using the Tsyganenko 04 storm time model \citep{tsyganenko2005modeling}. The first adiabatic invariant, $\mu$, corresponds to electron gyrational motion about the magnetic field; $K$, which is a combination of the first two adiabatic invariants, corresponds to bounce motion along the field lines; it is independent of particle mass and charge and is only invariant when field-aligned electric fields are absent; and $L^*$, which is commonly referred to as the Roederer $L$ related to the third adiabatic invariant \citep{roederer2012dynamics}, corresponds to drift motion around Earth.

In recent decades, the study of space weather has gained increasing prominence due to its profound impact on satellite functionality, communication systems, and terrestrial infrastructure. Among the myriad space weather phenomena, HILDCAA events have emerged as a critical driver of magnetospheric dynamics. These events, characterized by prolonged periods of elevated auroral activity, interact with the Van Allen radiation belts, regions of trapped energetic particles that play a vital role in shielding the Earth from cosmic radiation. The studies \citep{hajra2024ultra, da2019contribution, nema2024impact} discussed their impact on the outer radiation belt. These studies have found the enhanced flux and coexistence of  ULF wave activity using superposed epoch analysis of many HILDCAA events.

Although past research has explored the general influence of HILDCAA events on these belts, a comparative analysis and impact focusing on the duration of these events remains to be investigated. This study aims to fill this gap by investigating how the shortest and longest HILDCAA events differentially affected the structure and dynamics of the outer radiation belts of the Earth. We show that event duration is a key factor influencing particle acceleration and loss processes within the belts, with significant implications for both space weather forecasting and the design of resilient space systems. 

The manuscript is organized as follows: Section 2 outlines the methodological approach, Section 3 presents the comparative analysis results, and Section 4 discusses the implications of our findings and possible directions for future research.

\section{Materials and Methods}
The 1‐min resolution OMNI data set \url{https://omniweb.gsfc.nasa.gov/html/about_data.html} was used for solar wind and interplanetary magnetic field (IMF) parameters and Dst and AE geomagnetic indices \citep{rostoker1972geomagnetic, akasofu1981relationships}, where Dst index quantifies the globally averaged changes in the horizontal component of Earth's magnetic field at the magnetic equator. It is primarily influenced by the ring current, which intensifies during geomagnetic storms. Whereas, the AE index quantifies the currents flowing in the high-latitude ionosphere associated with auroral phenomena. It is derived from the horizontal magnetic field components recorded at multiple observatories located within the auroral zones. Higher AE values correspond to increased auroral activity and geomagnetic disturbances \citep{davis1966auroral}. The HILDCAA events have been identified by following the criteria given by \citet{tsurutani1987cause}. The near-Earth solar wind conditions during HILDCAA have been studied using the data available at \url{https://omniweb.gsfc.nasa.gov/}. The OMNIWeb data have already been time-shifted to the bow shock's nose, and the measured quantities in the GSM coordinate system have been used for further analysis \citep{russell1971geophysical, hapgood1992space}. The GSM coordinate system \citep{laundal2017magnetic} is particularly advantageous for studies focusing on the orientation and effects of the interplanetary magnetic field (IMF) on Earth's magnetosphere. By aligning the X-axis with the Earth-Sun line and incorporating the geomagnetic dipole in the X-Z plane, GSM coordinates facilitate the examination of solar wind-magnetosphere interactions, especially the solar wind-magnetosphere coupling processlike magnetic reconnection and geomagnetic storms.

For the radiation belt particle measurements, primarily we have used the observations from the twin spacecraft Van Allen Probes \citep{mauk2014science}, which are in proximity to the Earth's equatorial plane (\url{https://rbspgway.jhuapl.edu/data_instrumentationSOC}). For the current analysis, we have used only RBSP-A spacecraft data since the present study investigates longer time-scale processes, and both spacecraft are expected to see a similar environment. The relativistic electron and proton fluxes in differential energy channels have been obtained from the Relativistic Electron Proton Telescope (REPT) \citep{baker2014relativistic, baker2021relativistic} instrument and Magnetic Electron Ion Spectrometer (MagEIS), \citep{blake2014magnetic, claudepierre2015background, fennell2015van} aboard Van Allen Probes with latest version level 2 and 3. With 12 energy channels and 17 pitch angles, the REPT instrument provides highly resolved electron observations in terms of energy and pitch angle. For this study, the relativistic electron flux data corresponding to these HILDCAA events have been sourced from the Van Allen Probe Science Gateway, available at \url{https://rbspgway.jhuapl.edu/}. The Electric and Magnetic Field Instrument Suite and Integrated Science (EMFISIS) instrument onboard the Van Allen Probes recorded high-resolution electric and magnetic field data\citep{kletzing2013electric}, which was used to infer wave activity during the studied events.

\section{Observations} 
\subsection{Geomagnetic parameters' variation during two events}
\begin{figure}
    \begin{subfigure}[b]{0.35\textwidth}
        \centering
        \caption{}
        \includegraphics[width=3.0in, height=6.5in]{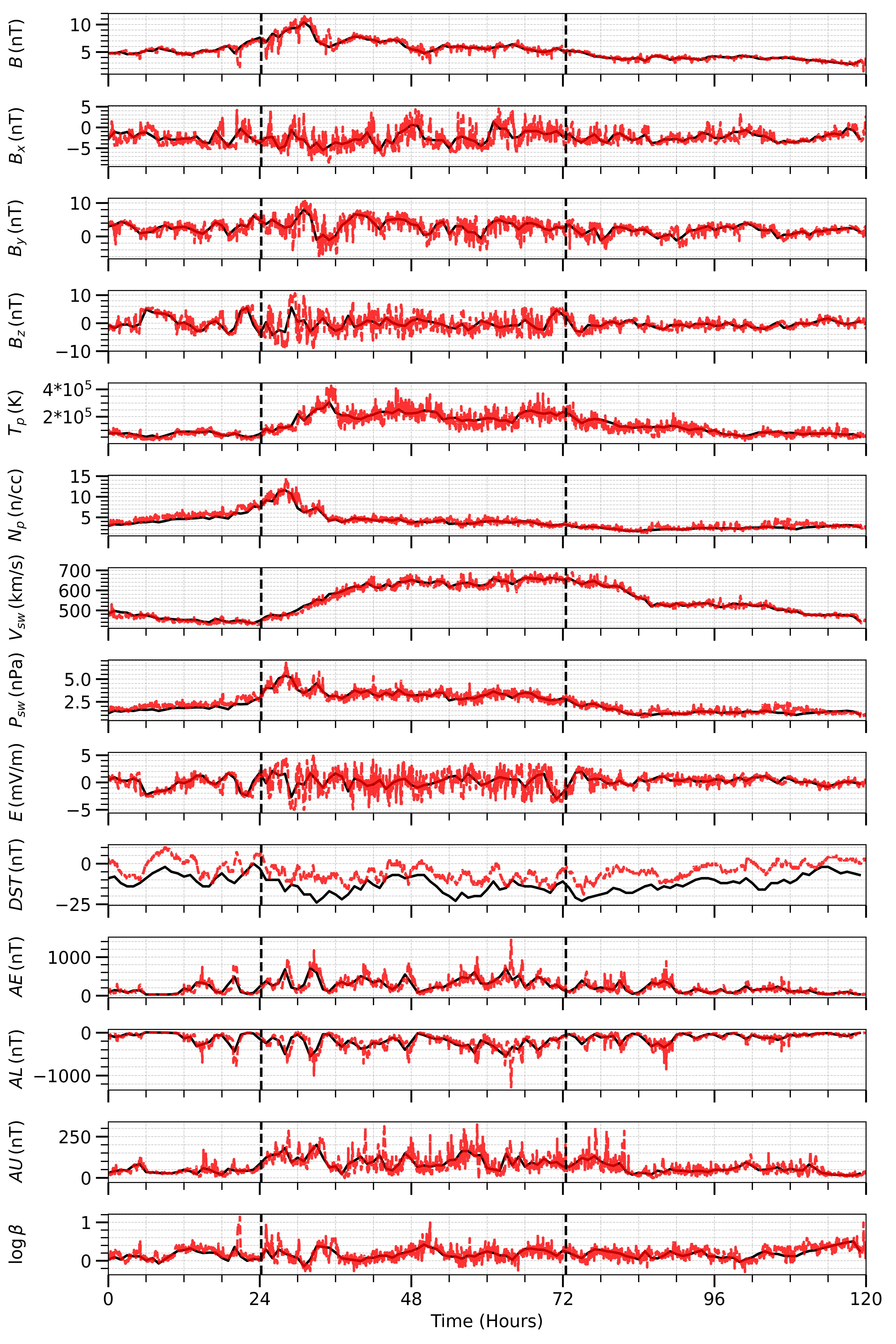}
        \label{fig:sub1.2}
    \end{subfigure}
    \hfill
    \begin{subfigure}[b]{0.45\textwidth}
        \centering
        \caption{}
        \includegraphics[width=3.0in, height=6.5in]{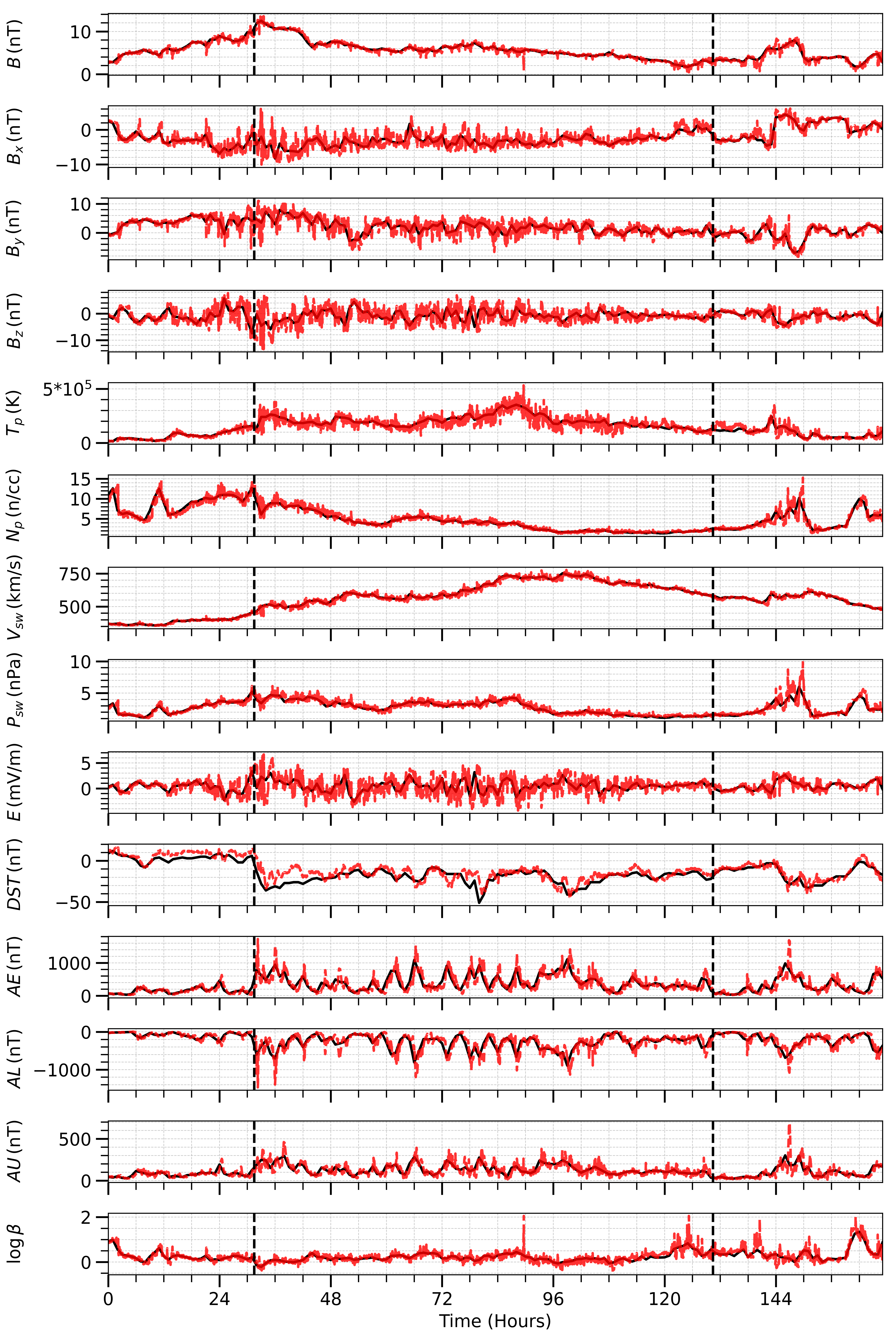}
        \label{fig:sub1.2}
    \end{subfigure}
    \caption{Interplanetary and geomagnetic parameters for shortest and longest HILDCAA events. The black dashed lines in the two plots indicate the start and end times of the two events. The shortest HILDCAA event started at 00:13 UT, December 10, 2015, and lasted till 00:28 UT, December 12, 2015. The longest HILDCAA event started at 07:28 UT, August 17, 2017, and lasted till 10:24 UT, August 21, 2017.}
    \label{fig:1}
\end{figure}
Fig.\ref{fig:1} represents the interplanetary parameters of both the HILDCAA events, together with one day before and after them. The first HILDCAA event observed from December 10 to 12, 2015, was the shortest event during the Van Allen Probes era. In contrast, the second HILDCAA event, which occurred from August 17 to 21, 2017, was the longest event of this period. The black line plots in Fig.\ref{fig:1} show 1-hour resolution, and the red line plots show 1-minute resolution. The two black dashed vertical lines in both plots represent the start and end times of the HILDCAA events. The two HILDCAA events shown in Fig.\ref{fig:1} follow the criteria defined by \citet{tsurutani1987cause}. The AE index (1 minute resolution) in both events never dropped below 200 nT for more than 2 hours. Both events are at least 2 days in length. Also, the events have peak AE values greater than 1000 nT. As shown in the Dst index plot, the events occurred outside the main phase of the geomagnetic storm or during the quiet period. In both events, IMF B\textsubscript{z} fluctuates during the period. The temperature and the solar wind velocity show enhancement after the onset in both cases.

\subsection{Temporal  variation of high and low energy Electron flux using REPT and MagEIS instruments}
\begin{figure}
    \centering
    \begin{subfigure}[b]{0.45\textwidth}
        \centering
        \caption{}
        \includegraphics[width=3.5in, height=6.0in]{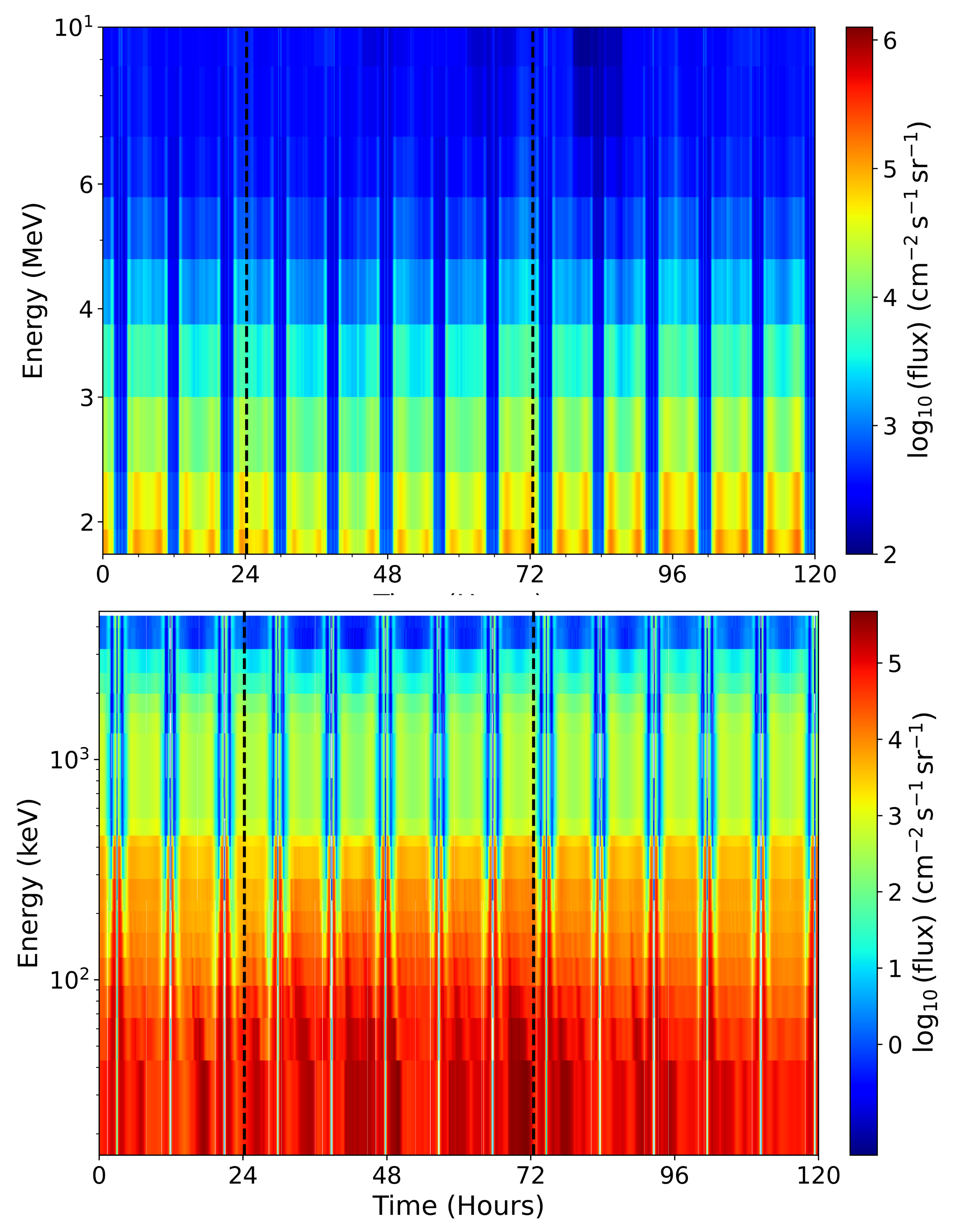}
        \label{fig:2.1}
    \end{subfigure}
    \hfill
    \begin{subfigure}[b]{0.45\textwidth}
        \centering
        \caption{}
        \includegraphics[width=3.5in, height=6.0in]{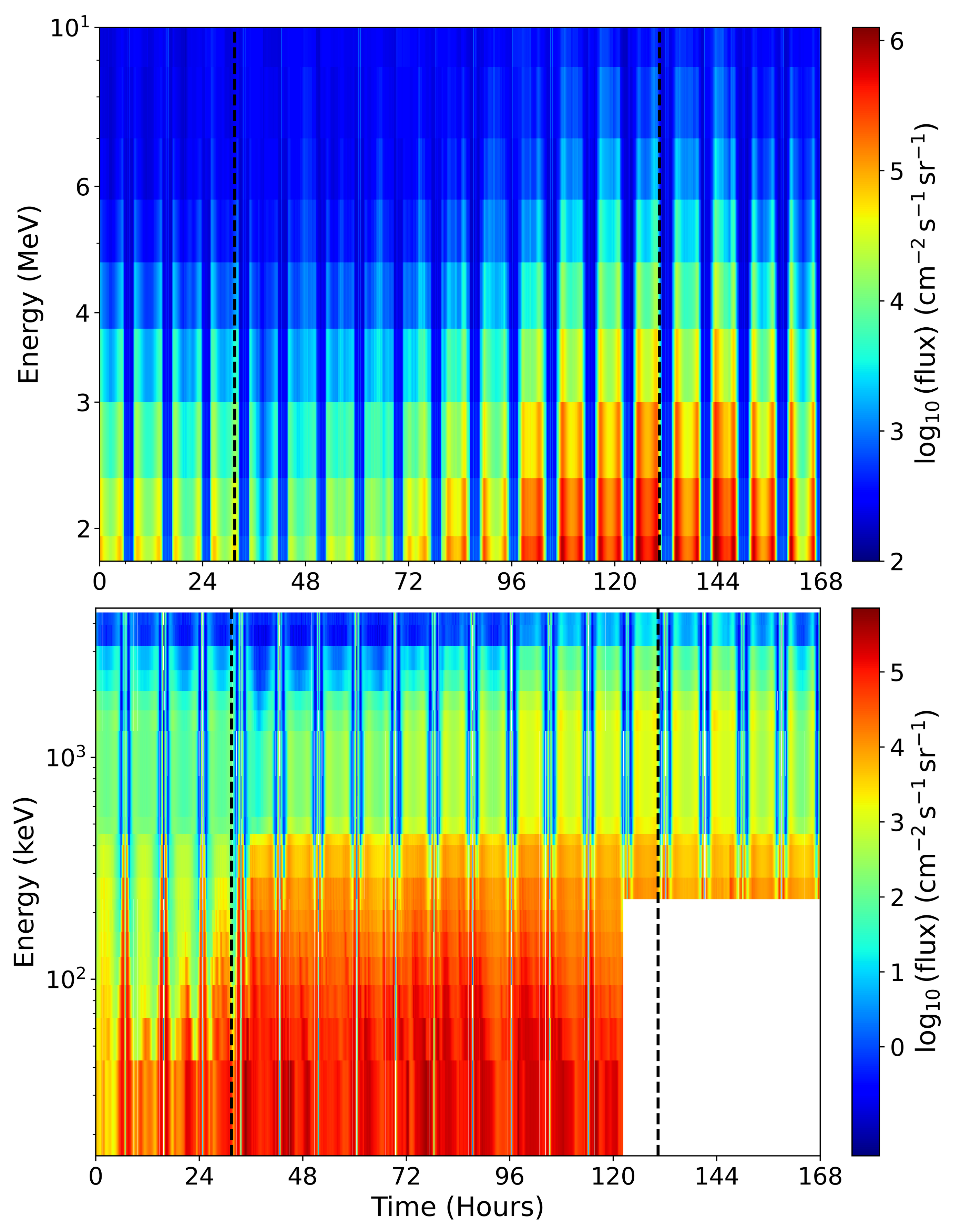}
        \label{fig:2.2}
    \end{subfigure}
    \caption{Variation of relativistic electrom flux for different energies using REPT and MagEIS in situ measurements for the shortest and longest HILDCAA event. The vertical dashed lines in all four plots show the periods of the HILDCAA event. The left plot is for the shortest HILDCAA event, and the right plot is for the longest HILDCAA event.}
    \label{fig:2}
\end{figure}

Figure \ref{fig:2} (top panels) illustrates the relativistic electron flux observed by REPT. Figure \ref{fig:2.1} (top-left panel) the flux values show a slight decrease at the onset of the event, followed by a modest increase after its conclusion. In contrast, in Figure \ref{fig:2.2} (top-right panel), the electron flux values were lower at the onset but increased after the event's midpoint. The highest electron flux was observed during the longest HILDCAA event in the REPT data, with an enhancement reaching up to 6 MeV.  The bottom panels of Figure \ref{fig:2} depict the flux variations of lower-energy electrons using MagEIS data. During the HILDCAA period, flux levels increased for both events. 

\subsection{Evolution of electron flux pitch angle distribution using REPT}
\begin{figure}
    \centering
    \begin{subfigure}[b]{0.45\textwidth}
        \centering
        \caption{}
        \includegraphics[width=2.5in, height=4.5in]{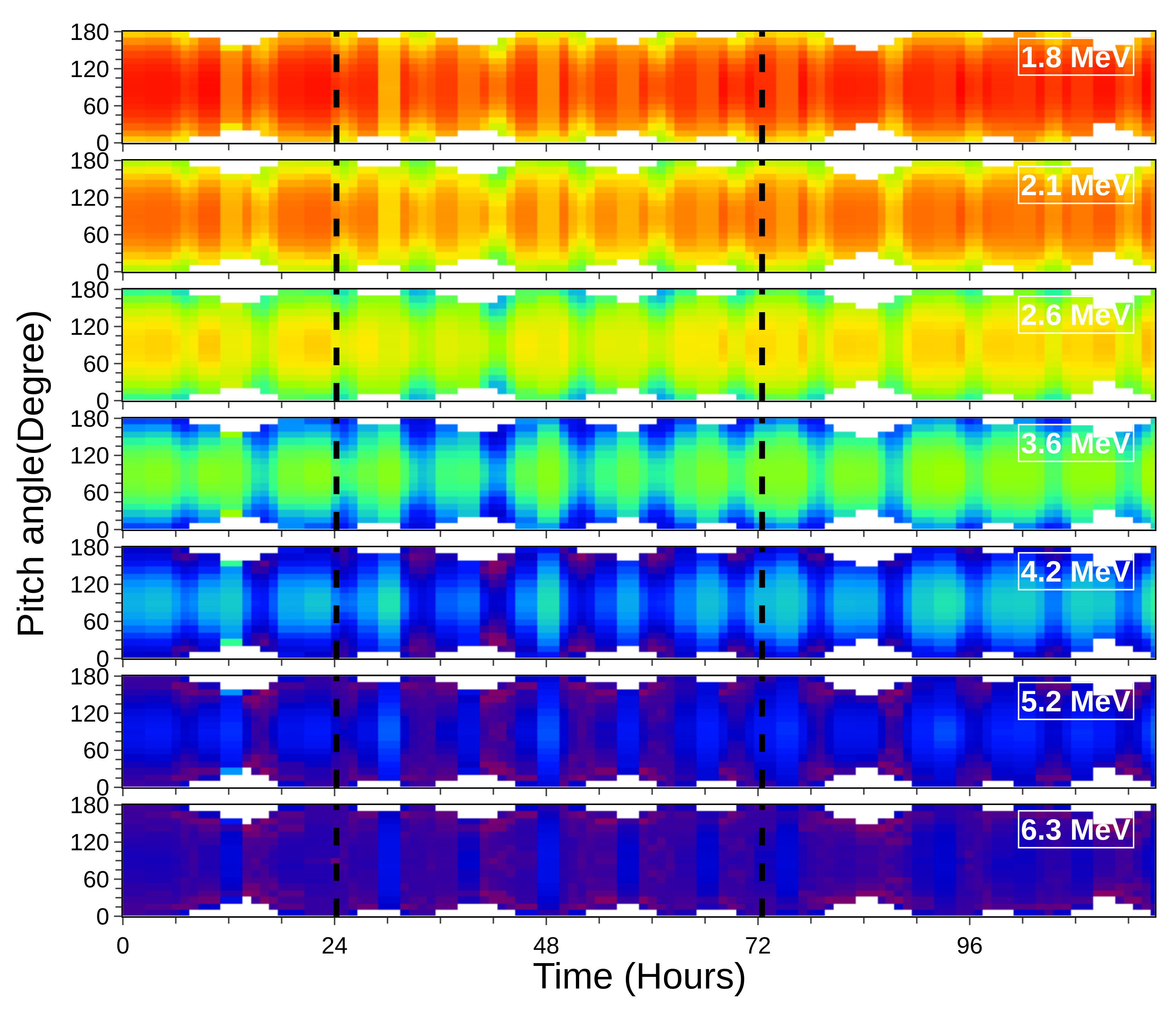}        
        \label{fig:3.1}
    \end{subfigure}
    \begin{subfigure}[b]{0.45\textwidth}
        \centering
         \caption{}
        \includegraphics[width=3.0in, height=4.5in]{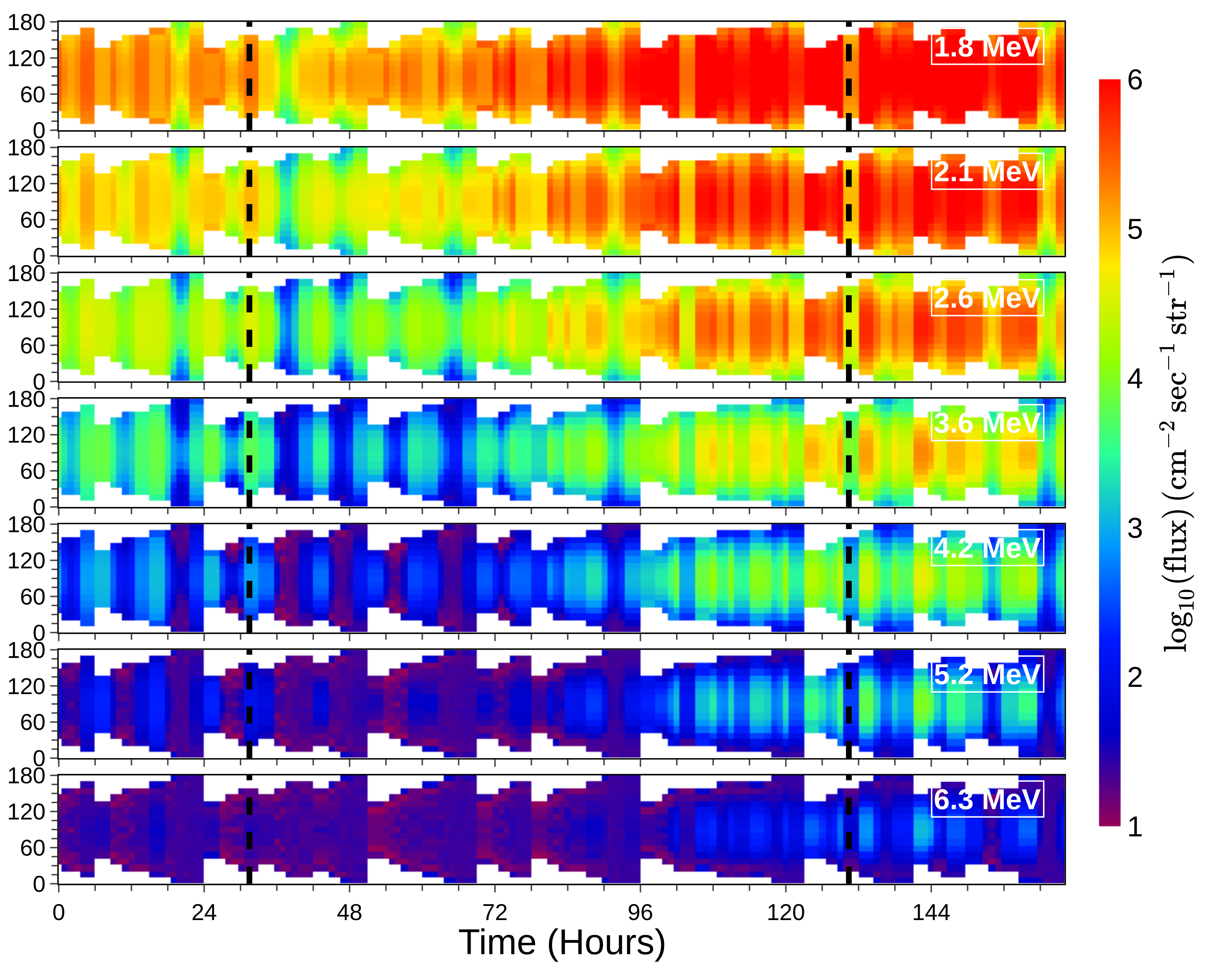}       
        \label{fig:3.2}
    \end{subfigure}
    \caption{Time evolution of pitch angle distribution of relativistic electron flux of the two events (shortest (a) and longest (b)). The vertical black dashed lines in the two plots show the durations of two HILDCAA events.}
    \label{fig:3}
\end{figure}

The relativistic electron flux variation with pitch angles has been shown in Fig.\ref{fig:3} for both events for seven energy channels (1.8-6.3 MeV). The black dashed vertical line shows the start and end of the HILDCAA event. This analysis provides an average representation of the variation of the relativistic electron flux during these events, to indicate particle injection during substorms for pitch angles. 

Figure \ref{fig:3} presents the time-resolved pitch-angle distributions of energetic electron fluxes, with the flux displayed on a logarithmic scale across multiple energy channels. In both figures, the color scale represents logarithmic electron flux, where warmer colors indicate higher flux values. The flux remains relatively higher at larger pitch angles (around 90$^\circ$) for both events.  Figure \ref{fig:3.1} exhibits a similar overall trend but with notable differences in flux distribution across pitch angles. In the second case, Figure \ref{fig:3.2}, the flux enhancement persists longer, reaching energies of approximately $\sim$6 MeV. Compared to Figure \ref{fig:3.1}, there is a more pronounced shift in flux toward higher pitch angles, potentially indicating enhanced scattering processes.  Additionally, the flux values display temporal variability across both energy and pitch angles. In the first case, relativistic electron flux initially decreases slightly before increasing, whereas in the second case, there is a significant decrease in relativistic electron flux. Moreover, the prolonged HILDCAA event clearly exhibits a delay in flux enhancements at higher energy channels compared to lower ones, indicating the presence of distinct time scales required for the acceleration process.

\subsection{Evolution of electron flux pitch angle distribution using MagEIS}
\begin{figure}
    \centering
    \begin{subfigure}[b]{0.45\textwidth}
        \centering
        \caption{}
        \includegraphics[width=2.5in, height=4.5in]{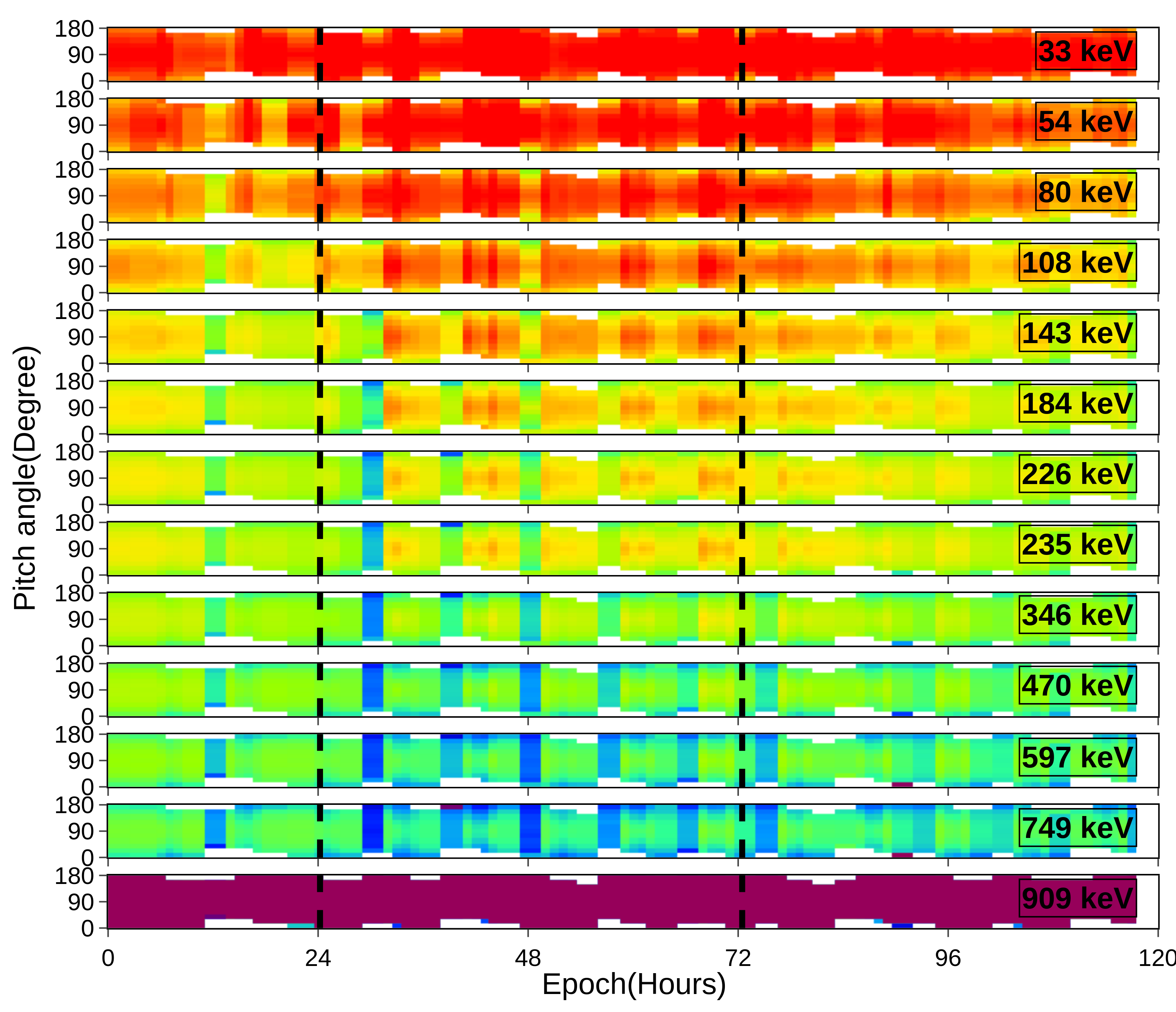}        
        \label{fig:4.1}
    \end{subfigure}
    \begin{subfigure}[b]{0.45\textwidth}
        \centering
        \caption{}
        \includegraphics[width=3.0in, height=4.5in]{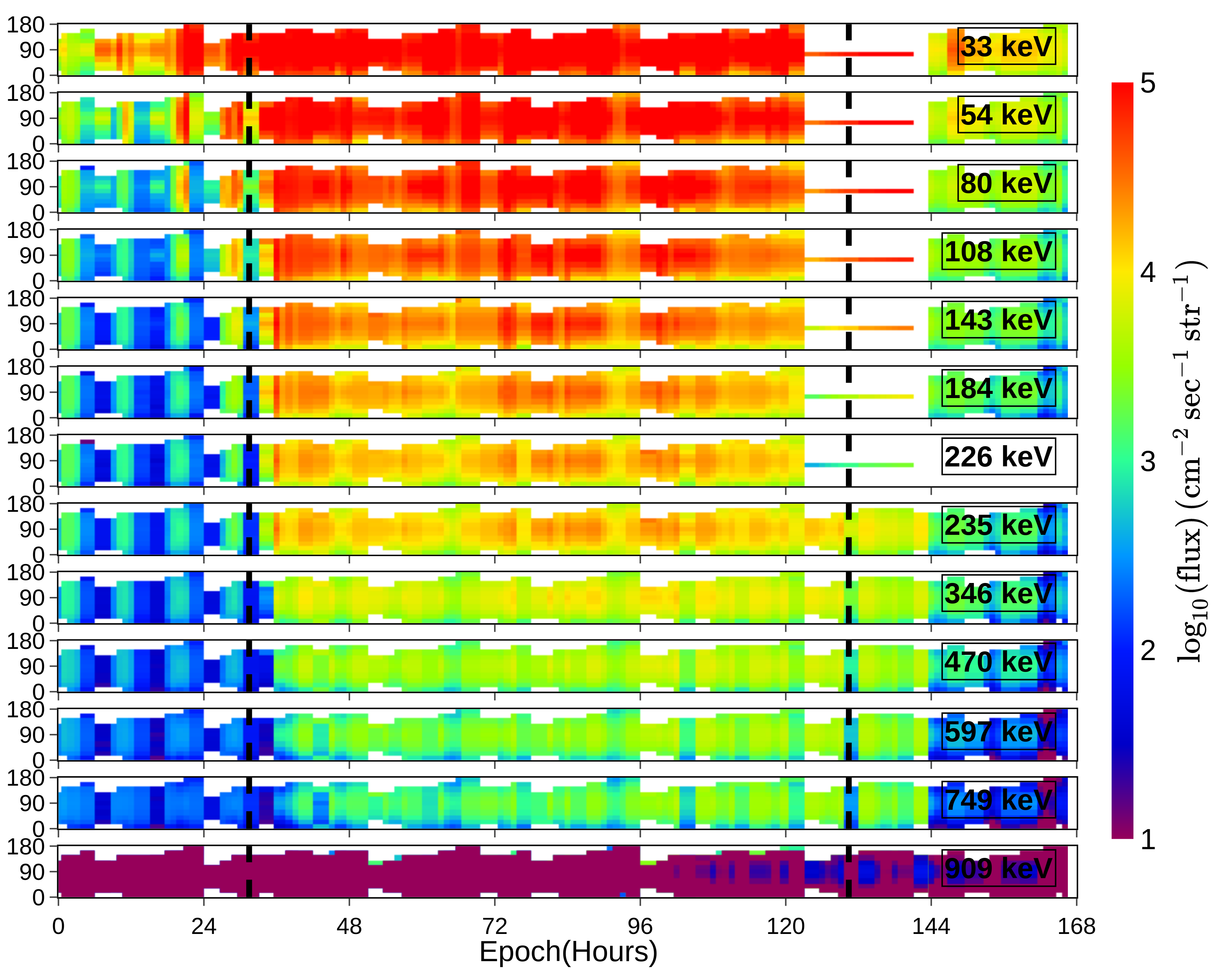}
        \label{fig:4.2}
    \end{subfigure}
    \caption{Pitch angle dependence of electron flux variation of the two events(shortest (a) and longest (b)). The vertical black dashed lines in the two plots show the start and end of the two HILDCAA events.}
    \label{fig:4}
\end{figure}

Figure \ref{fig:4.1} illustrates the variation in relativistic electron flux across different pitch angles using MagEIS data, with both plots displaying flux values up to 909 keV. The black dashed vertical lines in each plot indicate the start and end times of the short (a) and long (b) HILDCAA events.

In Figure \ref{fig:4.1}, the relativistic electron flux was comparatively higher before the onset of the event. For the shortest HILDCAA event, flux variations are observed up to approximately 346 keV. Notably, flux enhancement is more pronounced within a specific pitch-angle range of $\sim 60^\circ$ to $\sim 120^\circ$, corresponding to nearly trapped particles. In contrast, flux variations are lower for pitch angles around $\sim 0^\circ$ or $\sim 180^\circ$, which primarily consist of field-aligned particles. Whereas, Figure \ref{fig:4.2} illustrates the variation in relativistic electron flux during the longer HILDCAA event concerning pitch angles, using MagEIS data. In this case, flux variations extend to higher energy channels, reaching approximately 749 keV. Notably, flux enhancement becomes evident at the onset of the event. In most energy channels, this enhancement persists after the onset and continues until the end of the HILDCAA period. Similar to the shorter event, flux enhancement is more pronounced for trapped particles compared to field-aligned particles.

\subsection{Relativistic Electron flux variation across L-shells using REPT}

\begin{figure}
    \begin{subfigure}[b]{0.40\textwidth}
        \centering
        \caption{}
        \includegraphics[width=3.2in, height=6.0in]{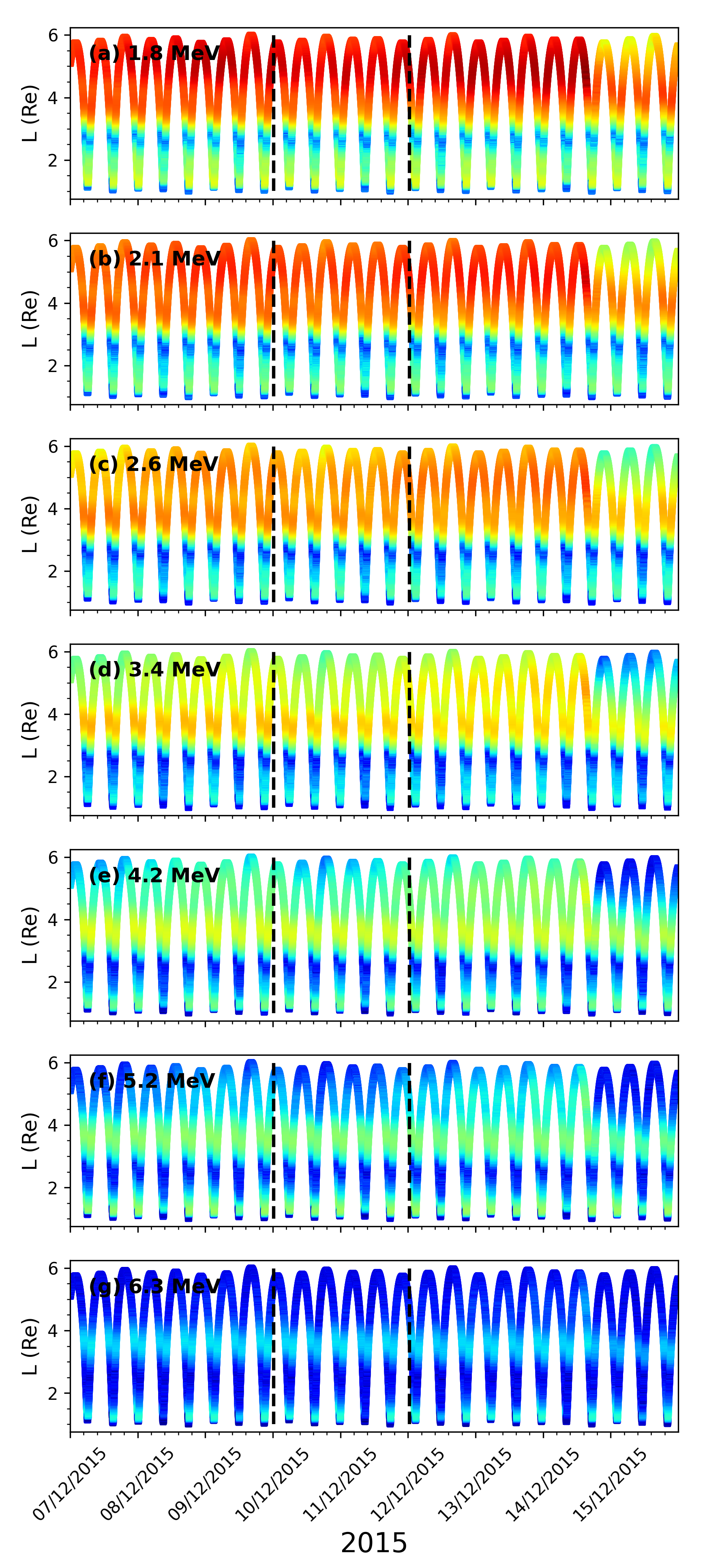}
        \label{fig:sub5.1}
    \end{subfigure}
    \hspace{25mm}
    \begin{subfigure}[b]{0.40\textwidth}
        \centering
        \caption{}
        \includegraphics[width=3.5in, height=6.0in]{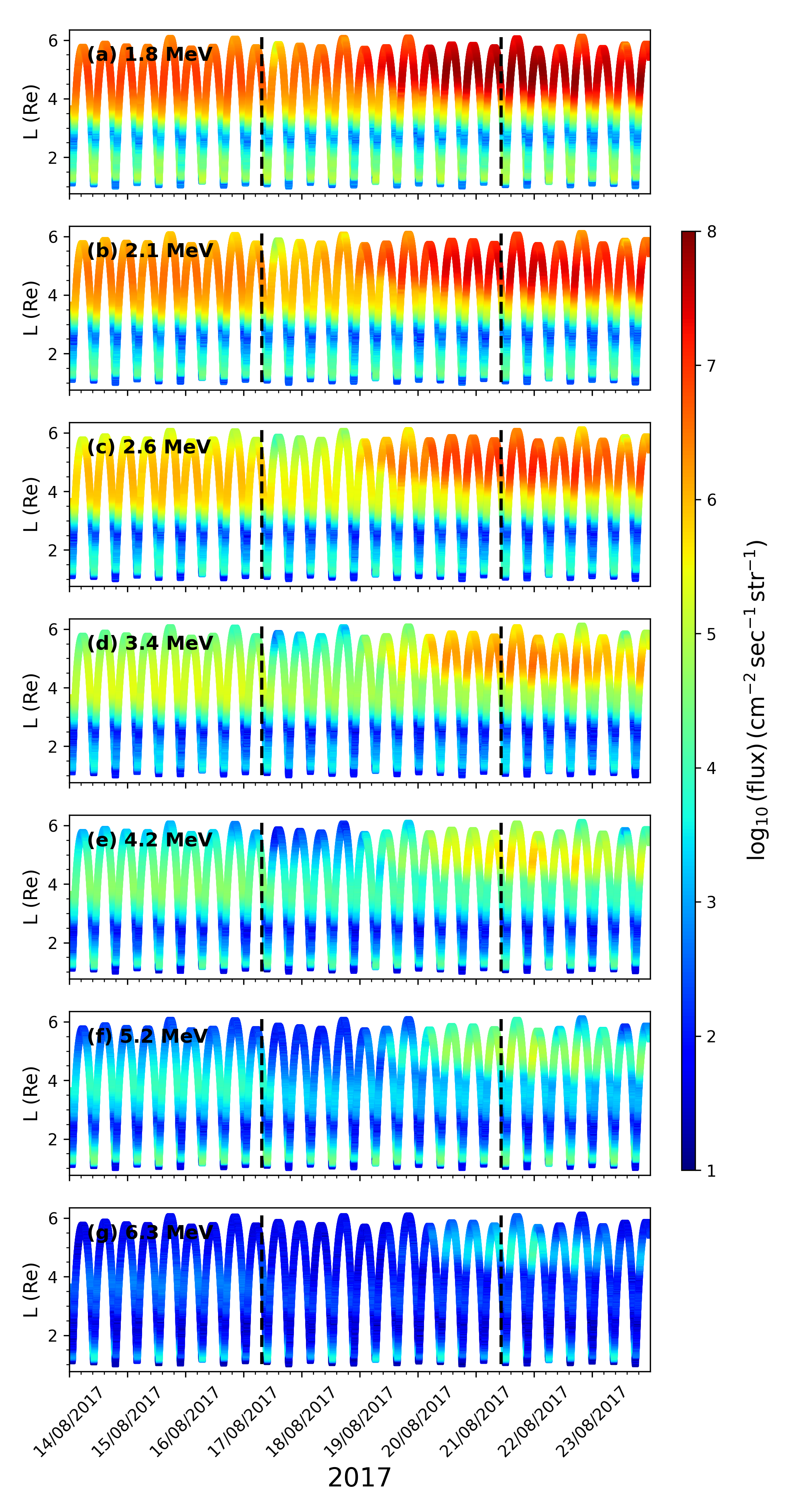}
        \label{fig:sub5.2}
    \end{subfigure}
    \caption{Fluxes of relativistic electrons observed by REPT with energies of $\sim$1.8–6.3 MeV. (a) Short (b) Long duration HILDCAA.}
    \label{fig:5}
\end{figure}

Fig.~\ref{fig:5} presents the electron flux variation during two HILDCAA events, as observed by the REPT instrument aboard the Van Allen Probes. The outer radiation belt fluxes show a highly dynamic nature across L-shells. The two panels, Fig.~\ref{fig:sub5.1} and Fig.~\ref{fig:sub5.2}, depict the electron flux variations for the shortest and longest HILDCAA events, respectively. For the short event, electron flux enhancement is dominant up to 5.2 MeV, beyond which variations become negligible. Flux variations are more pronounced above L = 4, where the electron flux initially increases before slightly decreasing after the onset. This trend remains consistent across all energy channels up to 5.2 MeV. Figure~\ref{fig:sub5.2} illustrates the electron flux variation during the longest HILDCAA event, which follows a similar pattern to the shorter event but with flux variations extending up to 6.3 MeV. In this case, the electron flux initially rises, decreases, which is prominent in higher L-shells and then increases again as compared to the event before event onset. In both events, the flux exhibits clear orbital modulation across L-shells ranging from L = 3 to 6, with peak intensities typically observed at higher L-shells (L $\geq$ 3). However, the 2017 event (longest event) shows a consistently greater flux enhancement compared to 2015 (shortest event), particularly at L = 4–6, where flux levels are significantly higher. Additionally, the longest event exhibits more pronounced temporal variability, characterized by stronger fluctuations and more frequent high-flux occurrences.  


\subsection{Electron flux variation across L-shells using MagEIS}

\begin{figure}
    \centering
    \begin{subfigure}[b]{0.45\textwidth}
        \centering
        \caption{}
        \includegraphics[width=3.0in, height=6.0in]{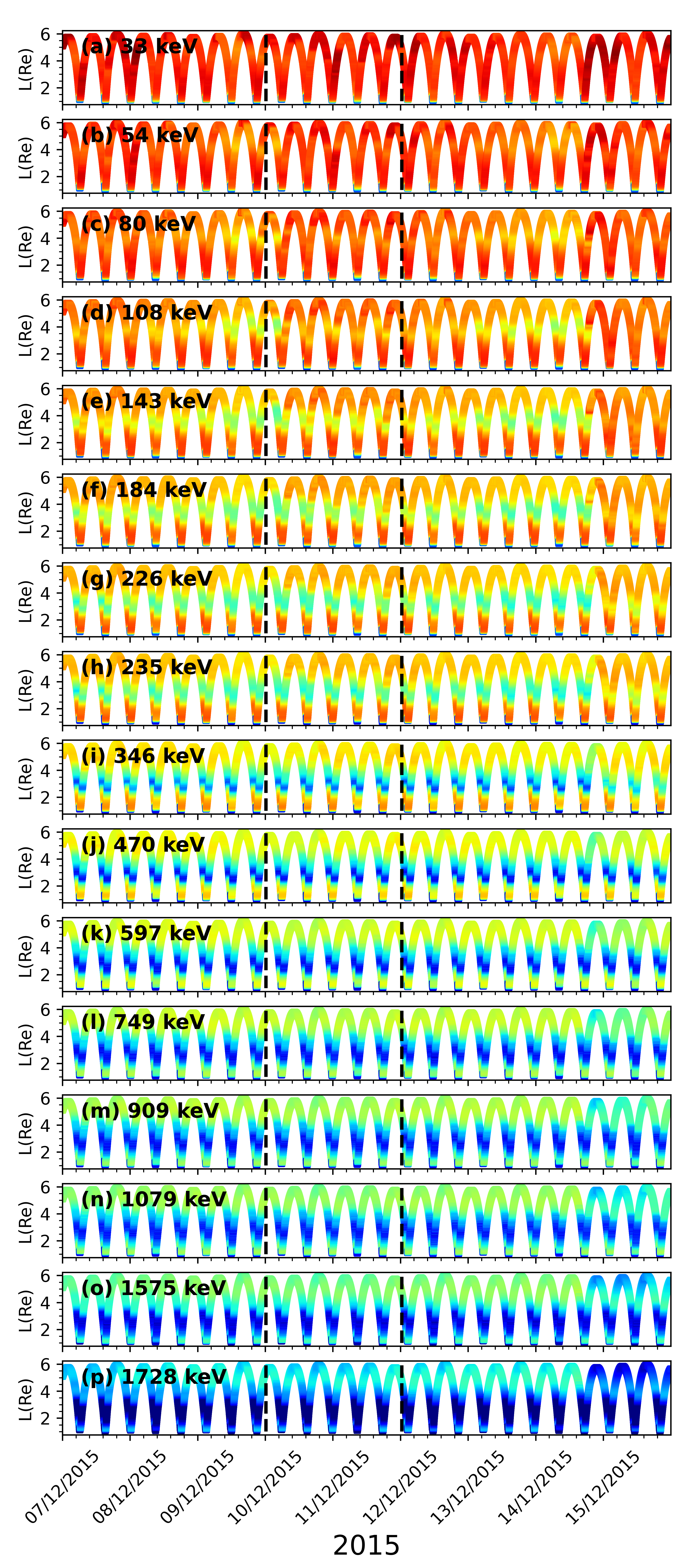}        
        \label{fig:6.1}
    \end{subfigure}
    \hfill

    \begin{subfigure}[b]{0.45\textwidth}
        \centering
        \caption{}
        \includegraphics[width=3.3in, height=6.0in]{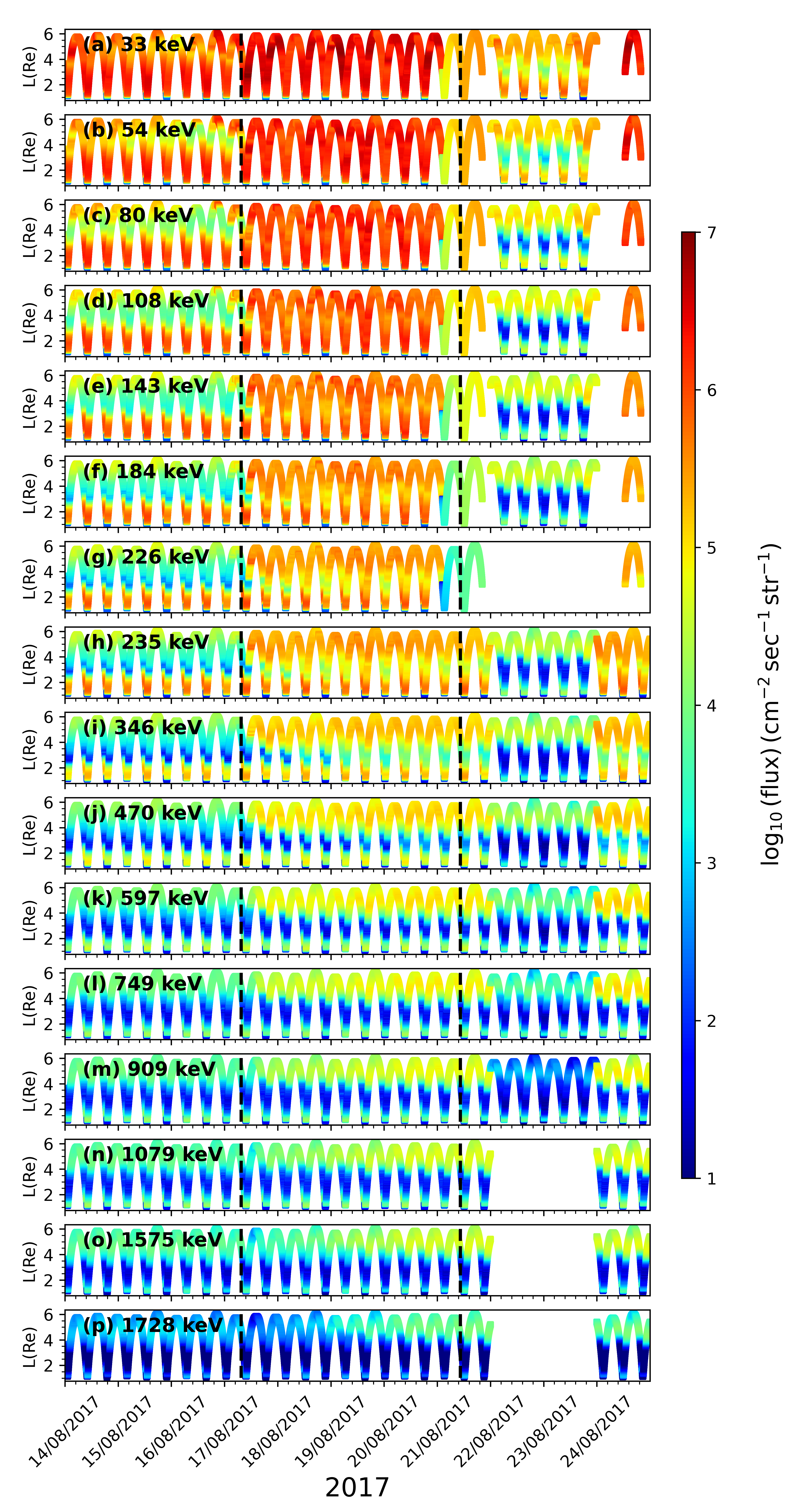}
        \label{fig:6.2}
    \end{subfigure}
    \caption{Fluxes of relativistic electrons using MagEIS with energies of $\sim$33–1728 keV.}
    \label{fig:6}
\end{figure}

Figure \ref{fig:6} illustrates the electron flux variation for both the shortest and longest HILDCAA events, as observed by the MagEIS instrument. MagEIS primarily measures electrons from approximately 20 keV to over 4 MeV and protons from around 60 keV to 1.3 MeV. With energy channels spanning from tens to thousands of keV, MagEIS serves as a complement to REPT, which measures relativistic and ultra-relativistic electrons.  For the shortest event, as shown in Figure \ref{fig:6.1}, the electron flux at the lowest energy channel remained elevated before, during, and after the event, showing less distinct changes. In contrast, for the longest event, the flux variation exhibits more distinct characteristics. Before the onset, the flux decreases slightly, followed by an increase after the event begins. This pre-onset decrease could be attributed to an enhancement in solar wind pressure, as evident in Figure \ref{fig:1}. After the event concludes, the flux decreases again. This pattern is consistently observed across almost all energy channels. As depicted in Figure \ref{fig:6.2}, the longest event demonstrates a pronounced flux enhancement throughout the HILDCAA period, highlighting the extended impact of the event on electron flux dynamics.

\subsection{Phase space density analysis}
When translating in-situ particle flux measurements into phase space density (PSD), it is appropriate to represent PSD as a function of quantities which are conserved under specific circumstances like the three adiabatic invariants that constrain the electron motion: $\mu$, $K$, and Roederer's $L^*$ parameter (which is inversely proportional to the third adiabatic invariant; thus, it is also an invariant) \citep{roederer2012dynamics}. Here we show time-dependent PSD radial profiles, i.e., PSD as a function of $L^*$ for fixed $\mu$ and $K$. 

\begin{figure}
    \centering
    \begin{subfigure}[b]{0.45\textwidth}
        \centering
        \caption{Short duration}
        \includegraphics[width=3.0in, height=6.5in]{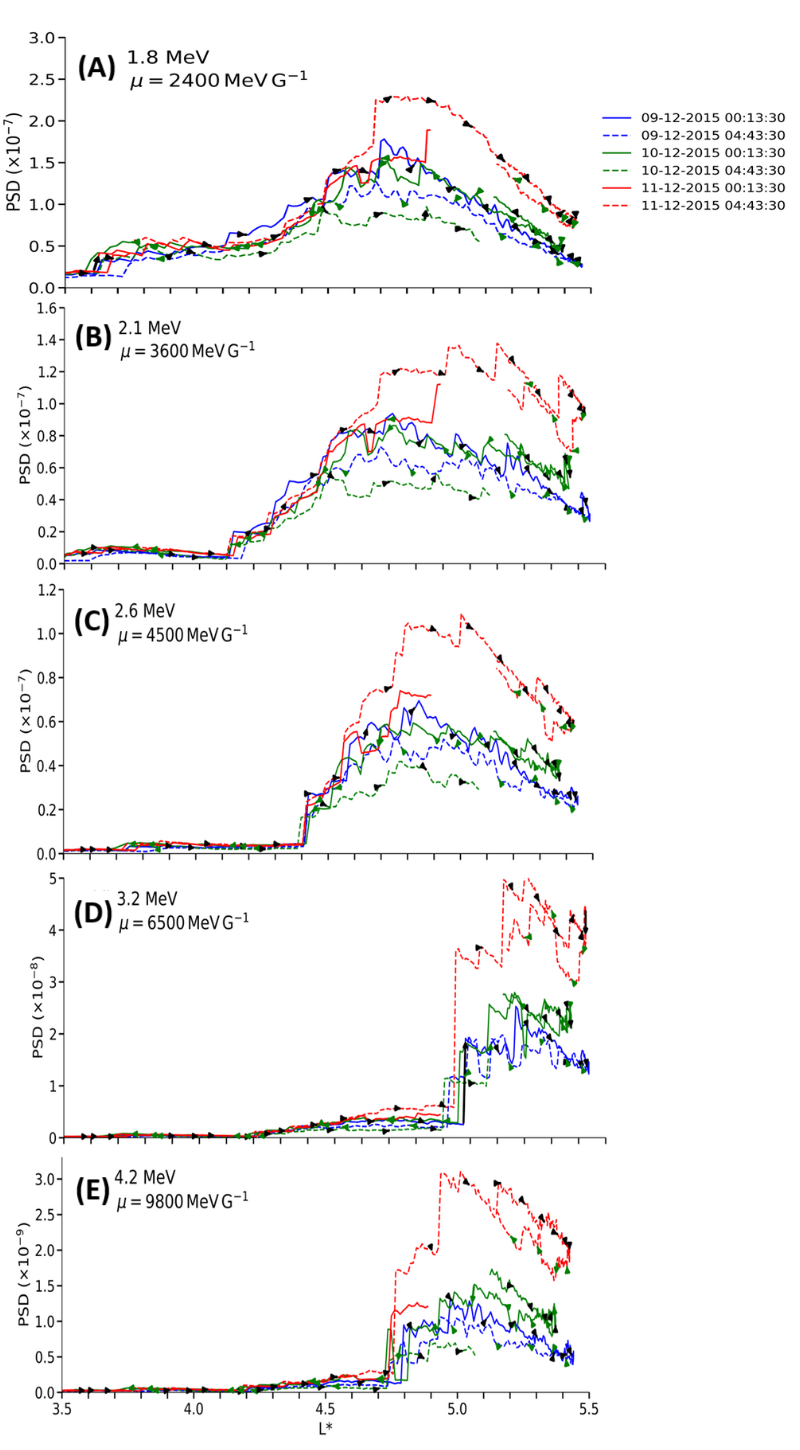}    
        \label{fig:7.1}
    \end{subfigure}
    \hfill
    \begin{subfigure}[b]{0.45\textwidth}
        \centering
        \caption{Long duration}
        \includegraphics[width=3.0in, height=6.5in]{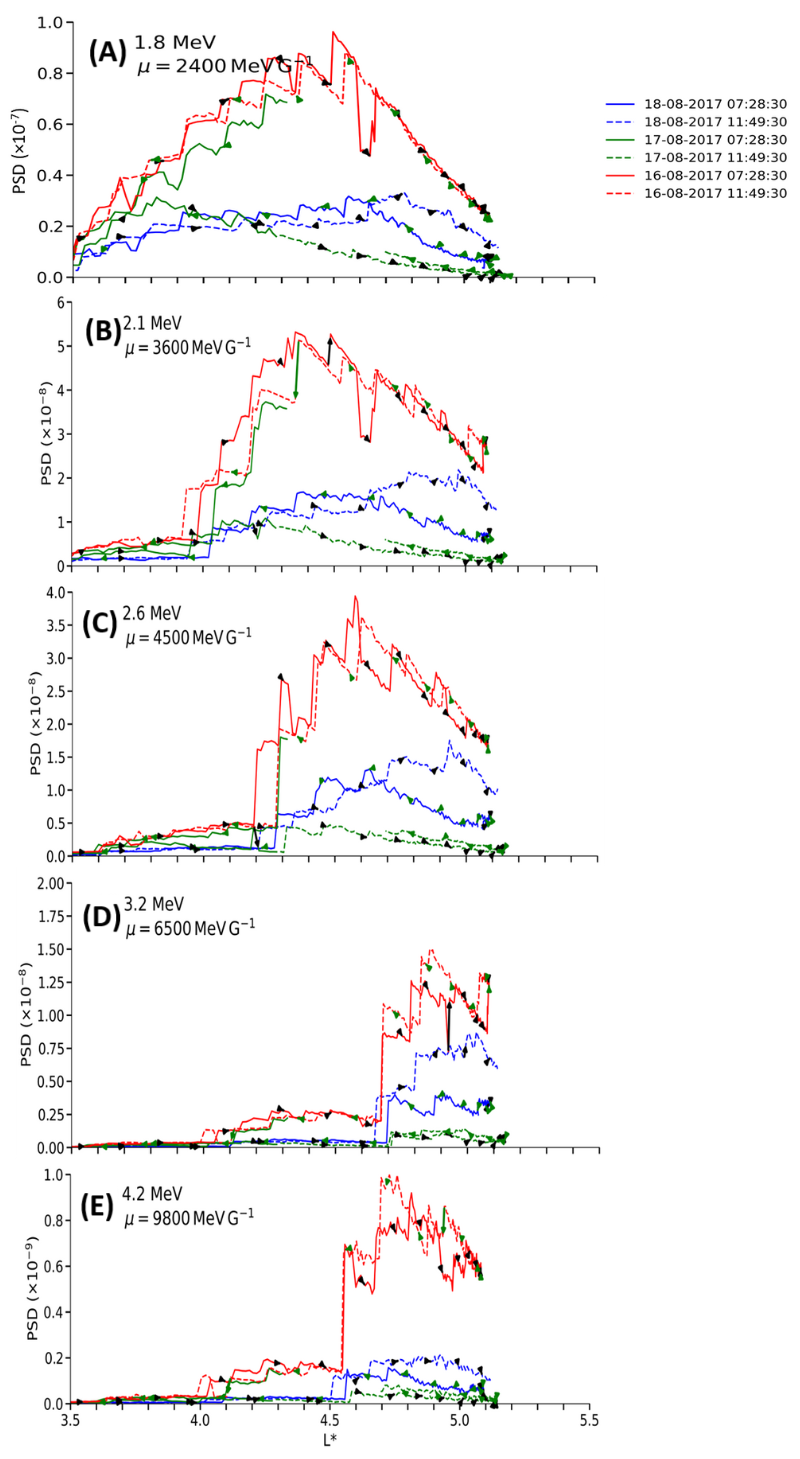}   
        \label{fig:7.2}
    \end{subfigure}
    \caption{Time evolution of phase space density radial profiles at fixed first ($\mu$) and second ($K$) adiabatic invariants for both events. The second invariant value is the same for Fig.\ref{fig:7.1}-\ref{fig:7.2}, while the first adiabatic invariant is varied, covering electron energy channels. The legends in Fig.\ref{fig:7.1}-\ref{fig:7.2} show the pre-onset (December 09, 2015; 00:13), onset (December 10, 2015; 00:13), and post-onset time (December 11, 2015; 00:13), with their 8.5 hours orbital period. In each panel, the solid line and dashed line represent data of the orbital motion of the spacecraft’s orbit, respectively, while blue, green, and red traces correspond to pre‐onset, onset, and post‐onset times.}
    \label{fig:7}
\end{figure}

Fig.\ref{fig:7}(\ref{fig:7.1} and \ref{fig:7.2}) shows PSD plots (in PSD $unit={c^3} {cm^{-3}} {MeV^{-3}}/PSDU$, c is the speed of light) of shorter and longer HILDCAA events encompassing the recovery of relativistic electron fluxes. PSD data for RBSP-A were obtained directly from \url{https://www.rbsp-ect.lanl.gov/data_pub/PSD/}, where PSD data can be found for a range of both $\mu$ and K values. For RBSP-A, values of $K = 0.11 G^{1/2} R_{\oplus}$ were used for various $\mu$. The adiabatic invariant, $\mu$ can be estimated using magnetic fields and energy channel values (in MeV). The estimated $\mu$ values can then be used to find the corresponding PSD values.

When analyzing the spatial evolution of PSD radial profiles in Fig.~\ref{fig:7} (\ref{fig:7.1}) and (\ref{fig:7.2}) during the inbound portions of the RBSP-A orbit, we observe that PSD initially increases and then decreases with decreasing L-shells. In the shortest HILDCAA event (Fig.~\ref{fig:7.1}), the PSD variations throughout one full orbit of the satellite are shown, covering the pre-onset, onset, and post-onset phases of the event.  The blue lines (dashed and solid) correspond to the pre-onset phase, the green lines (dashed and solid) represent the onset phase, and the red lines (dashed and solid) indicate the post-onset phase. The dashed lines denote the first half of the orbit, while the solid lines represent the second half. A key feature of the PSD profiles is the rising peak around \(L^* \sim 4.5\text{--}5.0\), which exhibits a negative gradient at higher \(L^*\) values for all \(\mu\) values. This negative gradient suggests that PSD decreases outward, indicating a potential local acceleration of electrons across all energy levels. Notably, this feature appears consistently across all \(\mu\) values, implying that the acceleration is not confined to a specific energy range.  

It can be noted that there is a shift in an increase in PSD values for increasing energies of electrons. For the lower energy channel 1.8 MeV, PSD increases at approximately $L^* = 3$. For energy channel 4.2 MeV, PSD starts to increase approximately $L^* = 5$. At the pre-onset of the HILDCAA event, the value of PSD peak decreases and shifts towards higher $L^*$. During the onset time of the HILDCAA event, that is, December 10, 2015, 00:13:30, PSD peaks shift toward higher $L^*$ shells. The peak values decrease for higher energy channels. For post-onset time, the peak PSD values increase as compared to other timelines. There is a shift in the peak PSD values towards higher $L^*$ shells. The same behavior of PSD is observed for a long HILDCAA event. But, the increase of PSD values occurs a little earlier when compared to a short event. It can be seen that the peaks in each PSD plot for each energy channel are higher for long HILDCAA as compared to shorter ones. The post-onset PSD values are consistently higher during both events. It can be due to sustained ULF/VLF wave-driven radial diffusion for a longer time.

In Fig.\ref{fig:7.1} at lower energies, such as 1.8 MeV (panel (A)), the PSD peaks can reach values as high as \(1.7 \times 10^{-7}\) PSDU for pre-onset conditions (blue line). This peak increases for post-onset (green line). As energy increases to 4.2 MeV (panel (E)), the peak values decrease to approximately \(1.5 \times 10^{-9}\) PSDU under pre-onset conditions (blue line) and \(3.0 \times 10^{-9}\) PSDU for post-onset at L=$\sim$4.5 to 5.5.  A decrease in PSD peak with L-shells suggests the electrons are being trapped, maybe due to their inward transport, and they can be locally accelerated, which prevents outward PSD growth \citep{iles2006phase}. A similar pattern is observed in panels (A, B, C, D and E) in Fig.\ref{fig:7.2}

\subsection{Spectral analysis}
\begin{figure}
    \centering
    \includegraphics[width=4.5in,height=4.5in]{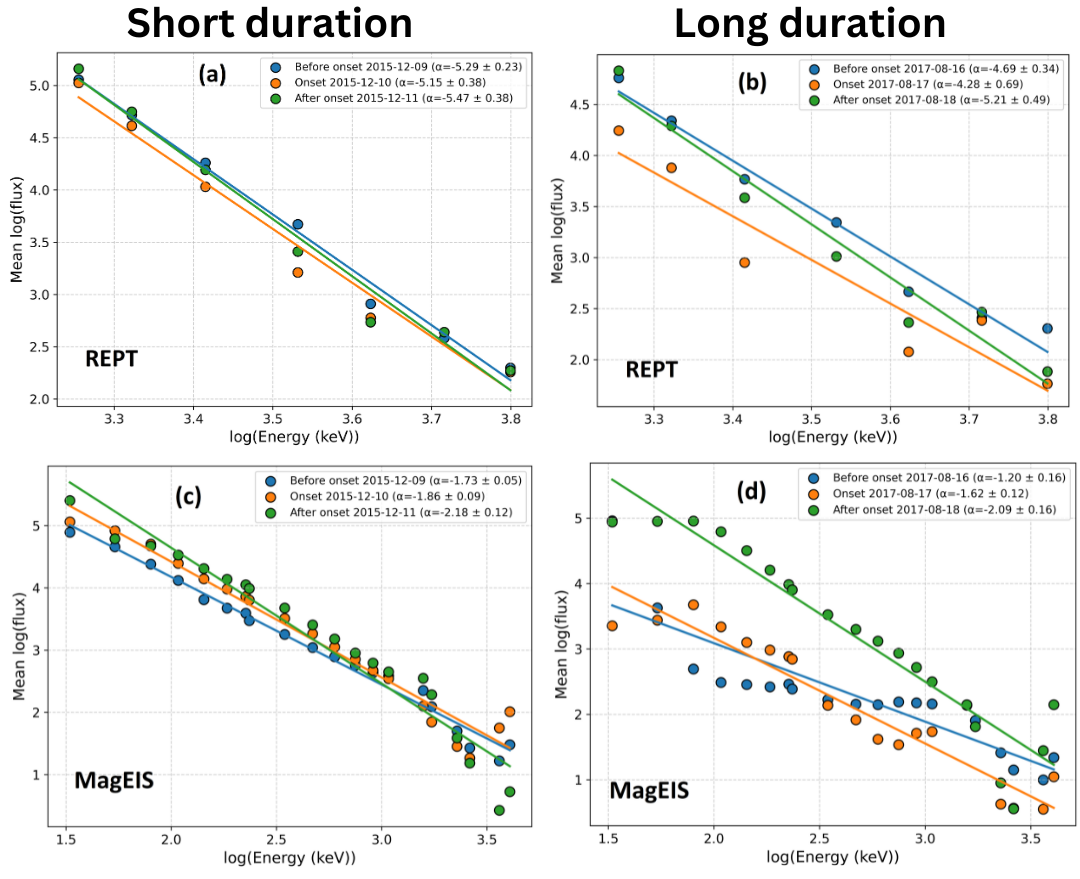}
    \caption{Spectral analysis of two HILDCAA events using REPT (top two) and MagEIS (bottom two). The spectral slope is indicated by parameter $\alpha$}
    \label{fig:8}
\end{figure}

Fig.\ref{fig:8} illustrates the log of mean relativistic electron flux as a function of log(Energy) for three different periods: before (2015-12-09, 2017-08-16), during (2015-12-10, 2017-08-17), and after (2015-12-11, 2017-08-18) the onset of the two HILDCAA events. The spectral slopes, represented by the parameter $\alpha$, indicate a progressive steepening that suggests an increased energy loss mechanism or a reduction in the population of particles of higher energy after onset. 

The spectral index steepening is more prominent in the first case, where mean log(flux) is taken with log(Energy), Fig.\ref{fig:8}(a) (top panel on left) as compared to Fig.\ref{fig:8}(b) (top panel on right, which shows longer duration of HILDAA), with a noticeable change from -5.29 ($\pm$0.23) to -5.47 ($\pm$0.38) before onset and after onset. In the second case (top (b) panel) of the longest event, the variation is relatively high (-4.69 to -5.21) with an error of $\pm$ 0.34 before the onset and $\pm$ 0.49 after the onset. These two panels are derived from REPT data. In second case Fig.\ref{fig:8}(c) and \ref{fig:8}(d) (bottom panels), mean log(flux) is compared with log(Energy) in all the three timelines (before onset, at onset and after onset) using MagEIS data with $\alpha$ for shorter event changes from -1.73 ($\pm$0.05) to -2.18 ($\pm$0.12). For a longer event, the difference in the $\alpha$ value is higher, changing from -1.20 ($\pm$0.16) to -2.09 ($\pm$0.16).

\subsection{HFR and WFR spectra analysis}
\begin{figure}
    \centering
    \begin{subfigure}[b]{0.45\textwidth}
        \centering
        \includegraphics[width=3.0in, height=3.0in]{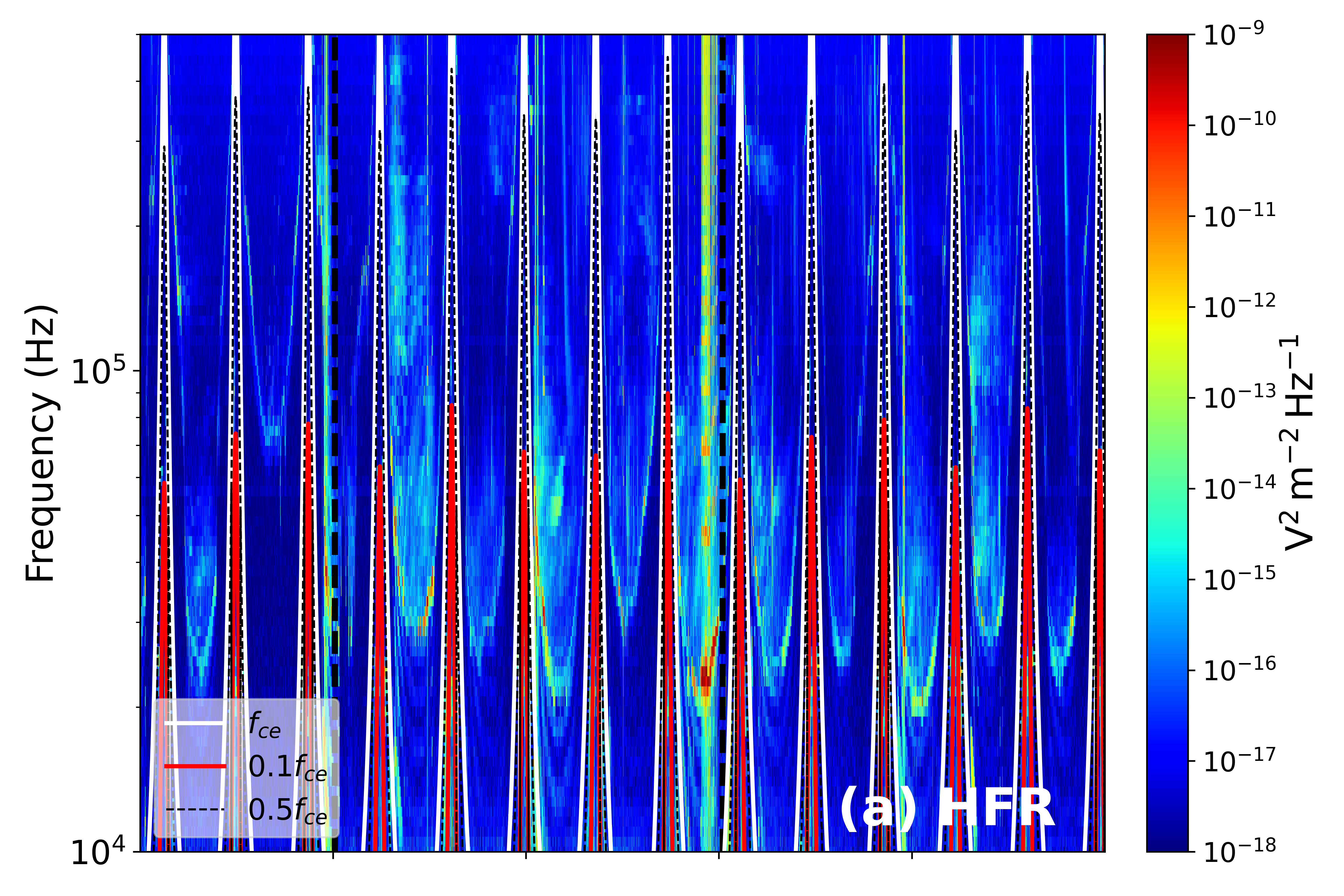}
        \label{fig:9.1}
    \end{subfigure}
    \hfill
    \begin{subfigure}[b]{0.45\textwidth}
        \centering
        \includegraphics[width=3.0in, height=3.0in]{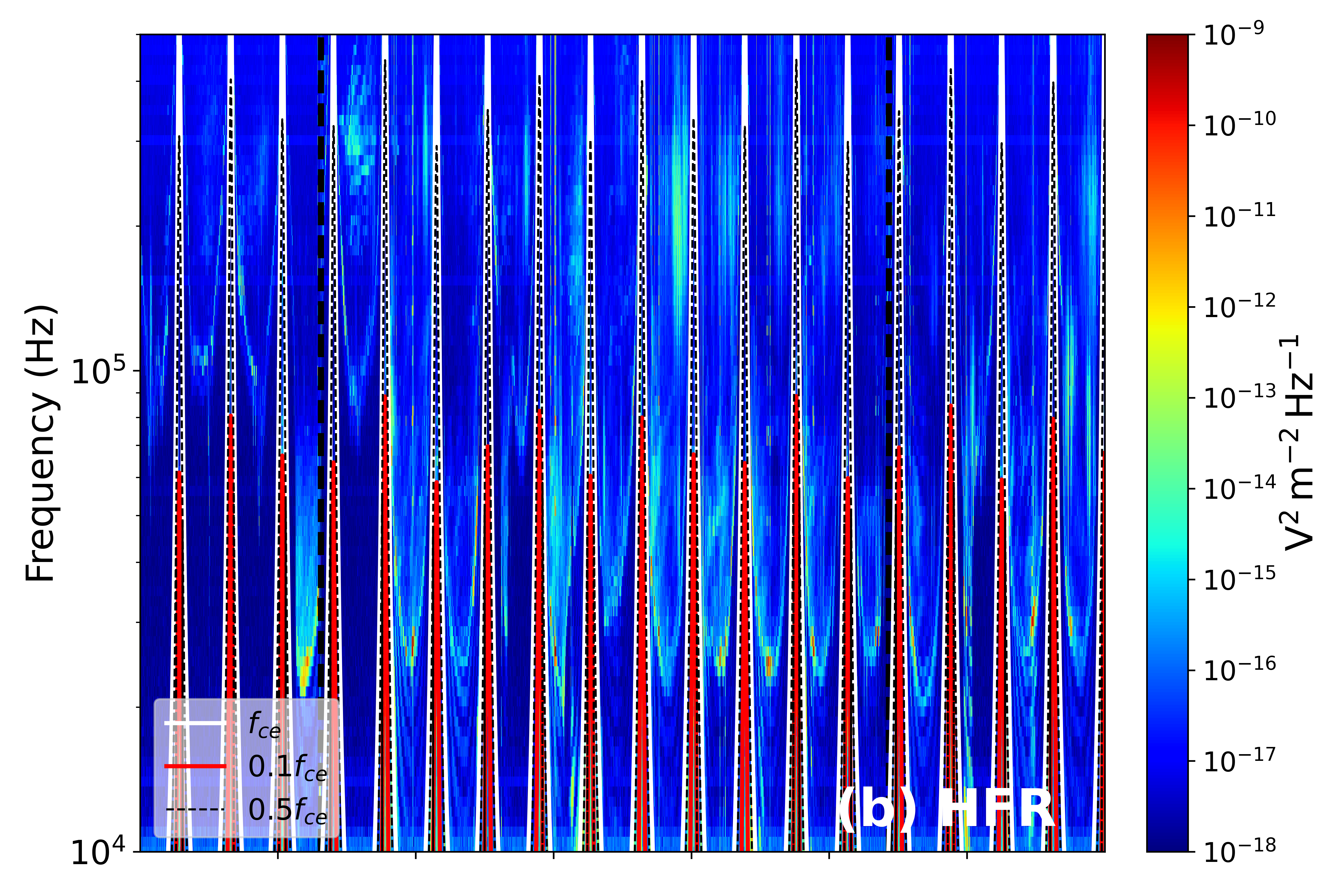}
        \label{fig:9.2}
    \end{subfigure}
    \vspace{0.1mm}
    \begin{subfigure}[b]{0.45\textwidth}
        \centering
        \includegraphics[width=3.0in, height=3.0in]{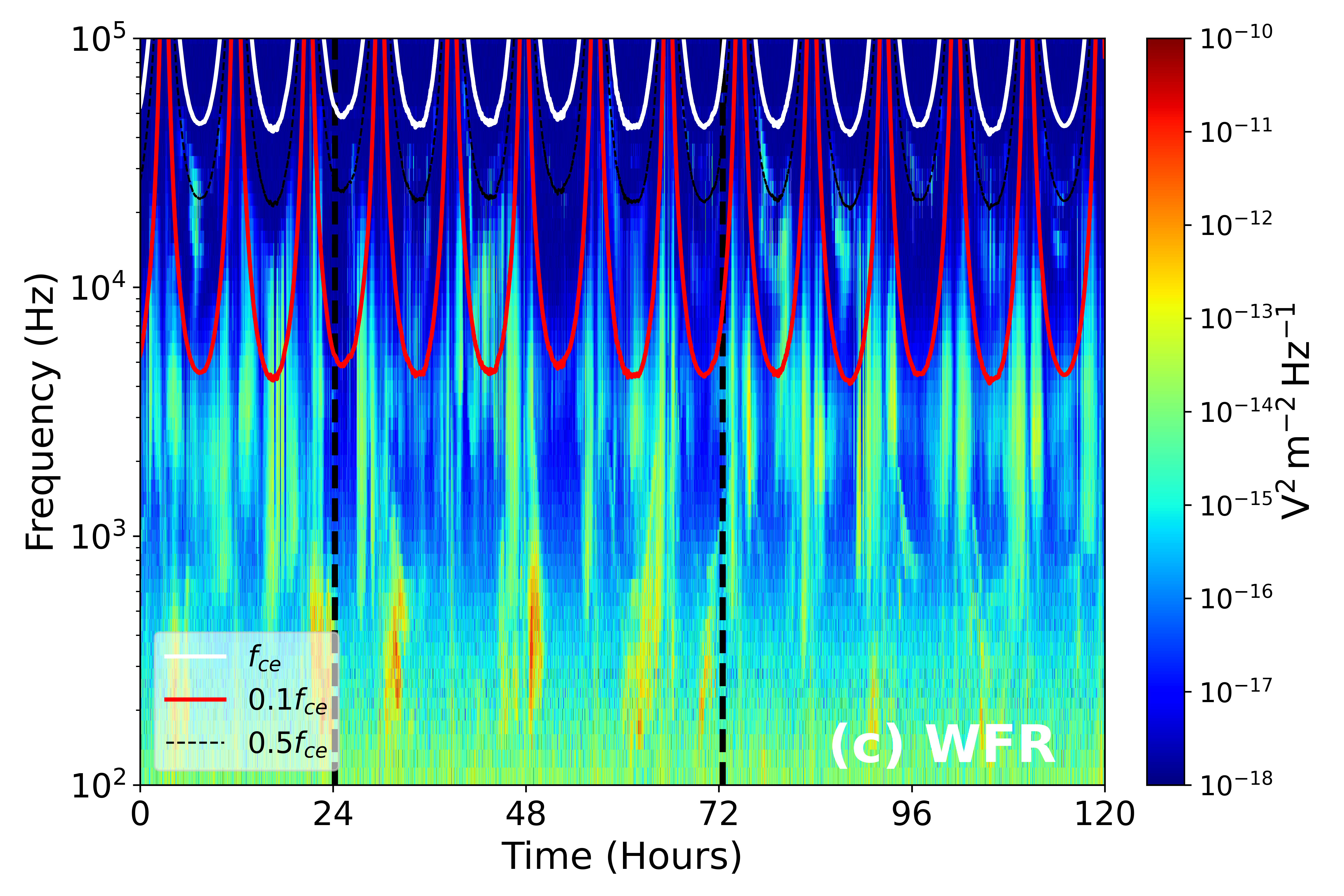}
        \label{fig:9.3}
    \end{subfigure}
    \hfill
    \begin{subfigure}[b]{0.45\textwidth}
        \centering
        \includegraphics[width=3.0in, height=3.0in]{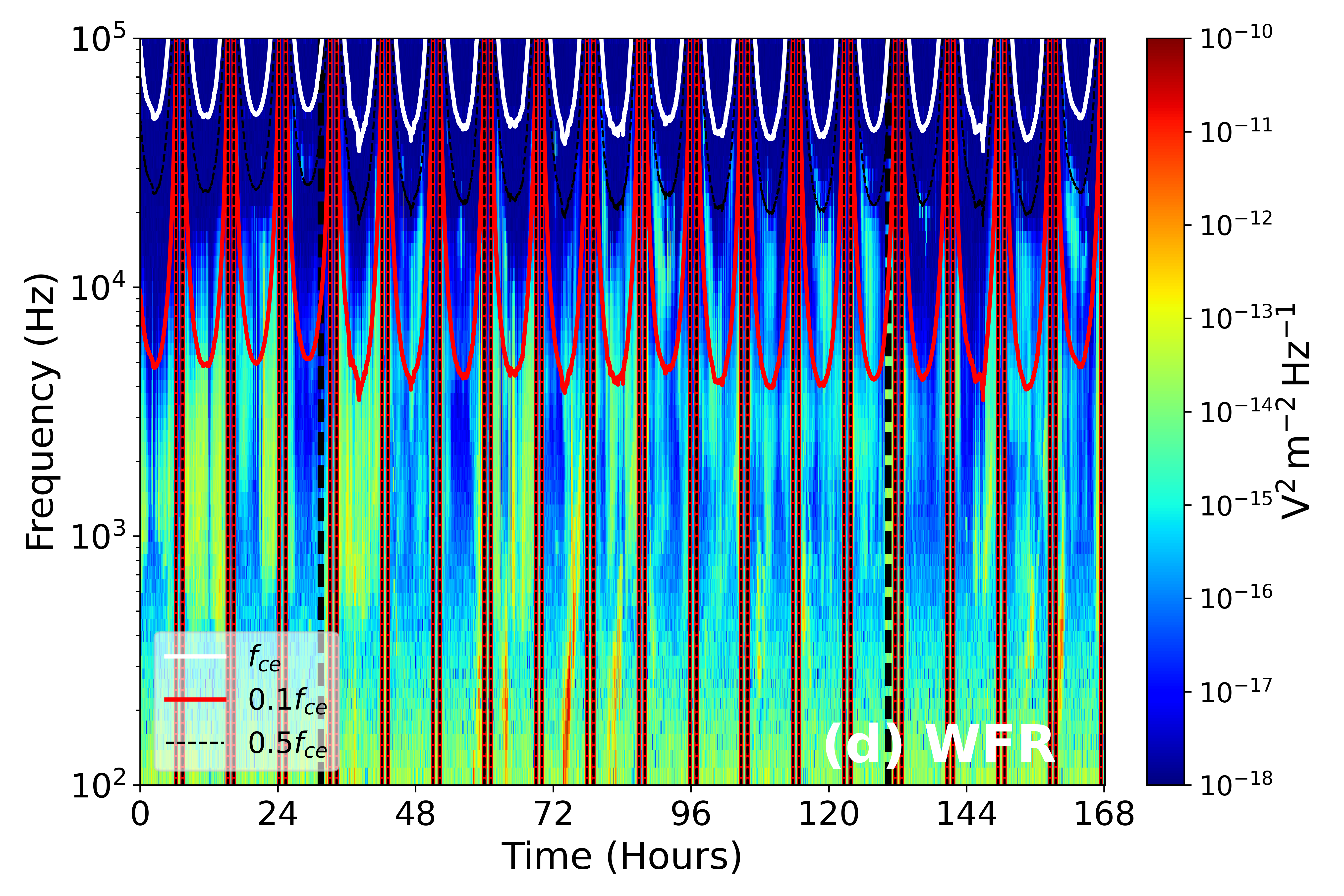}
        \label{fig:9.4}
    \end{subfigure}
    \caption{HFR and WFR spectral analysis for two HILDCAA analyses. The 2 vertical black dashed lines in each panel represent the start and end time of the HILDCAA event. The colorbar on the right side of each panel shows the wave power. Electron cyclotron frequency ($f_{ce}$) and harmonics ($0.5f_{ce}$ and $0.1f_{ce}$) are shown in white, dashed black lines, and red, respectively.}
    \label{fig:9}
\end{figure}

Fig.\ref{fig:9} represents high frequency (HFR) and waveform receiver (WFR) spectrum from the EMFISIS instrument of Van Allen Probe-A. The top two panels represent the wave spectrum from HFR, and the bottom two panels represent the wave spectrum using WFR for the shortest (a,c) and longest (b,d) HILDCAA event. The electron cyclotron frequency and its harmonics are overplotted over each spectrum.

While the right spectrum, Fig.\ref{fig:9}(b), covers a longer time of 168 hours, the left spectrum (Fig.\ref{fig:9}(a)) covers roughly 120 hours. Long-term trends become more apparent in the right spectrum due to the longer duration.  Nonetheless, both wave spectra show similar spectral features, with power enhancements in the same frequency ranges near $0.1f_{ce}$ band, suggesting sustained chorus wave activity during both the HILDCAA events.

In panels Fig.\ref{fig:9}(c) and \ref{fig:9}(d) WFR spectrogram is plotted for two time ranges using the EMFISIS WFR-spectral-matrix. The enhanced power indicated stronger chorus wave activity in the second HILDCAA event compared to the first one. This increased power in longer duration indicates possibly heightened wave-particle interactions, which might be responsible for increased high-energy electron fluxes in the second event. 

\subsection{ULF wave activity}
The duration of a HILDCAA event determines how long radiation belt electrons are exposed to enhanced geomagnetic activity. Longer HILDCAA and chorus wave activity are linked to higher maximum energies of accelerated relativistic electrons \citep{hajra2024ultra}. While chorus waves contribute to acceleration, we also examine ULF wave activity across different event durations. In short events, ULF waves may briefly boost electron energies and fluxes, whereas long events enable sustained wave-particle interactions, leading to prolonged acceleration or diffusion. Comparing ULF activity in short and long events helps clarify these effects.

\begin{figure}
    \centering
    \begin{subfigure}[b]{0.45\textwidth}
        \centering
        \includegraphics[width=3.0in, height=2.5in]{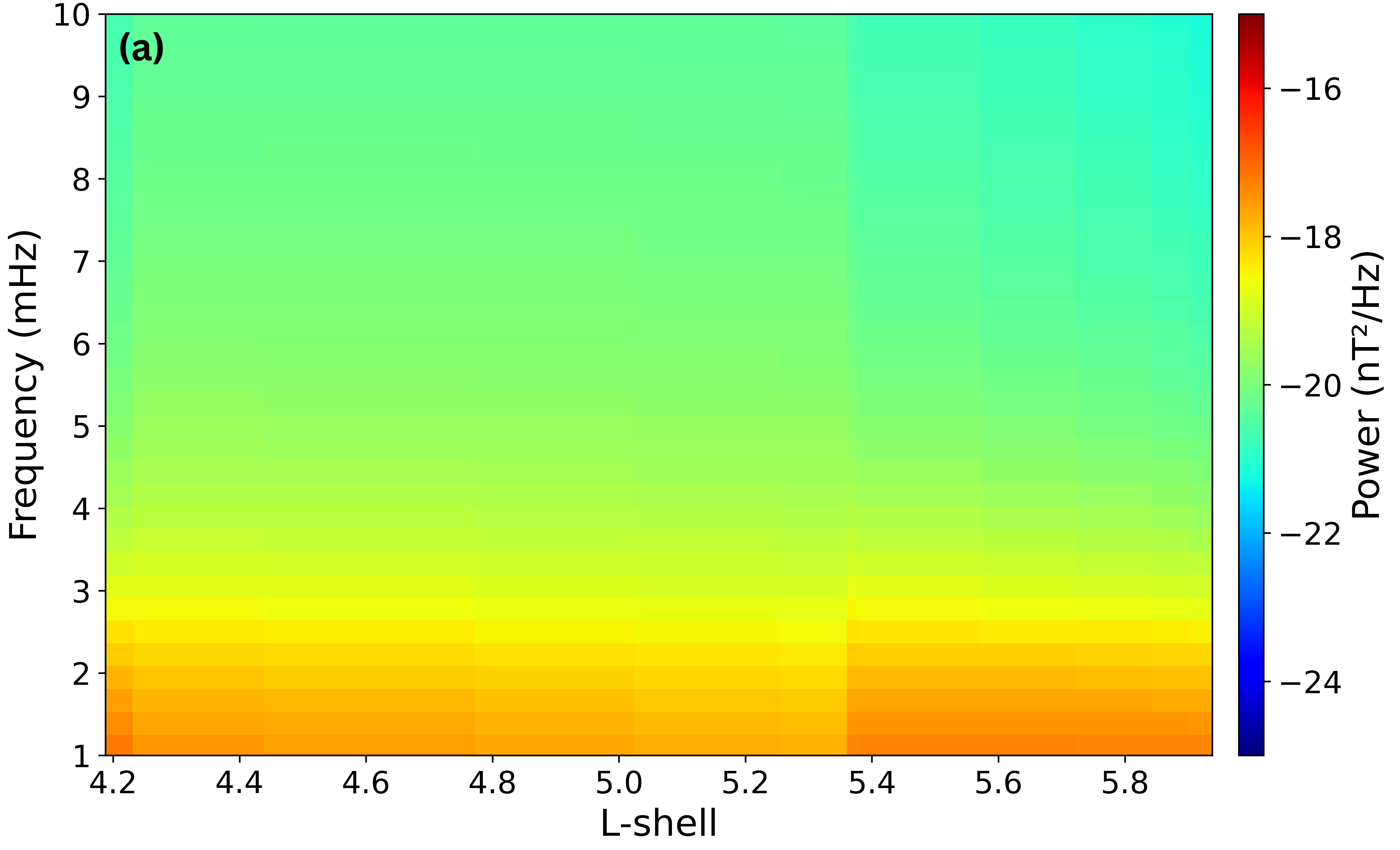}
        \label{fig:10.1}
    \end{subfigure}
    \hfill
    \begin{subfigure}[b]{0.45\textwidth}
        \centering
        \includegraphics[width=3.0in, height=2.5in]{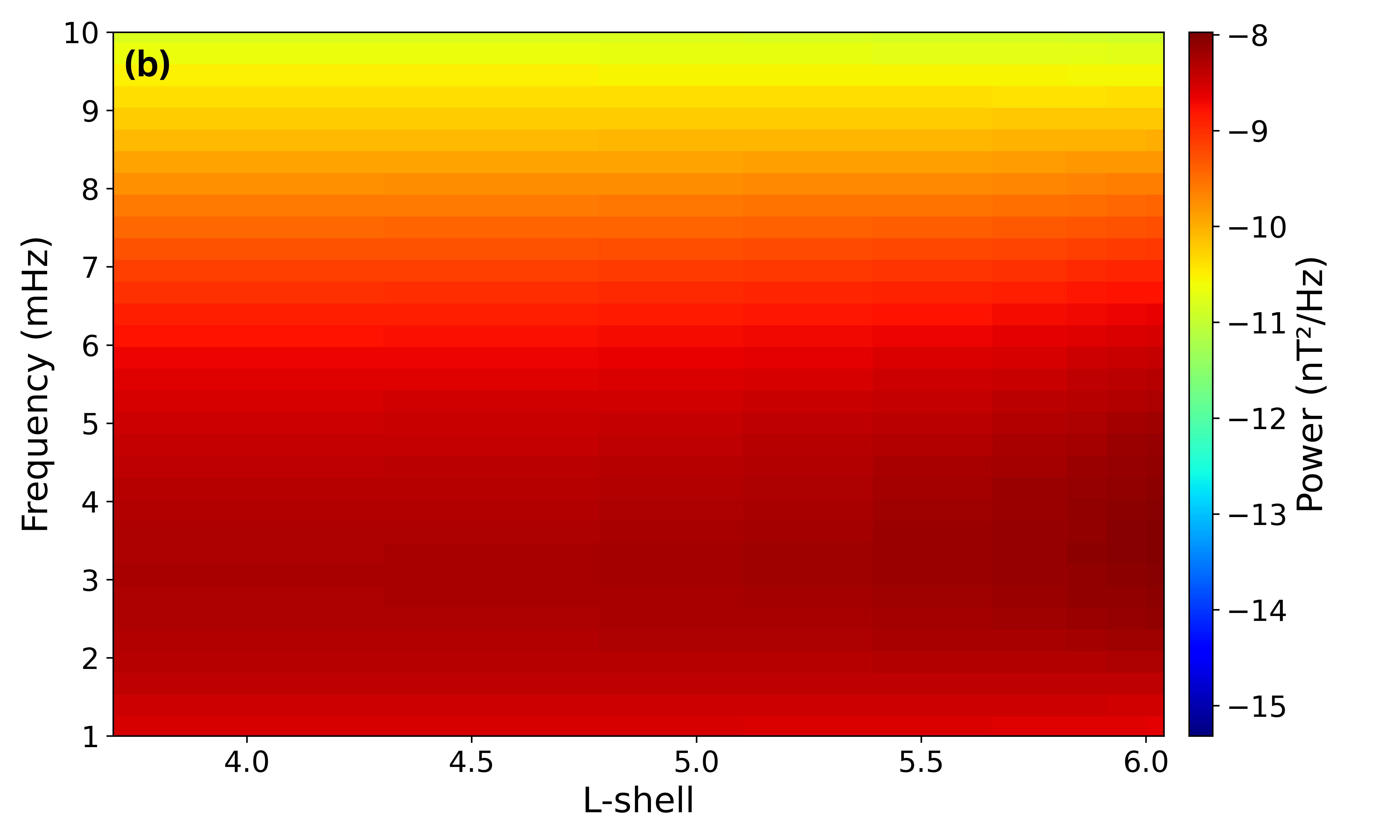}
        \label{fig:10.2}
    \end{subfigure}
    \caption{ULF wave analysis of the two HILDCAA events. The two plots derived from EMFISIS display the wave power variation with respect to L-shells and frequency.}
    \label{fig:10}
\end{figure}

We used magnetic field data from the EMISIS instrument onboard RBSPA. To identify ULF waves, we applied a Butterworth band-pass filter. Fig.\ref{fig:10}(a) and \ref{fig:10}(b) show two spectrograms of ULF wave power (in nT\textsuperscript{2}/Hz) plotted against frequency (1–10 mHz) and L‐shell ($\sim$4-6) for the shortest and longest HILDCAA events. In both panels, warmer colors correspond to higher power levels, revealing the spatial (L‐shell) and spectral (frequency) distribution of ULF oscillations. The left panel (Fig.\ref{fig:10}(a)) exhibits a lower maximum power range (approximately -16 to -24 nT\textsuperscript{2}/Hz), which corresponds to the short HILDCAA event. In contrast, the right panel (Fig.\ref{fig:10}(b)) shows a higher power range (roughly -8 to -15 nT\textsuperscript{2}/Hz) and displays more uniformly elevated intensities across the entire 1-10 mHz band, indicating a broader and more intense wave activity, corresponding to the longest HILDCAA event. These differences may reflect distinct driving mechanisms or magnetospheric conditions, such as changes in solar wind dynamic pressure, varying geomagnetic activity levels, or the location of the plasmapause, that can alter both the amplitude and spatial distribution of ULF waves in the outer magnetosphere.

\section{Discussion and Conclusions}

This study demonstrates that the duration of High-Intensity Long-Duration Continuous Auroral Electrojet Activity (HILDCAA) events plays a critical role in shaping the dynamics of Earth’s outer radiation belt. Using data from NASA’s Van Allen Probes, we analyzed two contrasting events—a short-duration and a long-duration HILDCAA—and observed that while both contribute to electron acceleration and flux enhancements, their effects differ significantly in magnitude and temporal evolution. Short-duration HILDCAA events induce rapid but transient increases in electron energy and flux, with enhancements that dissipate soon after the event ends. In contrast, long-duration events sustain elevated electron fluxes for extended periods, suggesting cumulative acceleration effects. Relativistic particles observed by REPT show an initial decrease after the onset of HILDCAAs, and then, with a finite time delay, an increase in fluxes. Whereas, low energy electrons observed by MagEIS show almost instantaneous enhancement after the onset of HILDCAA. This contrasting feature indicates a possible two-step acceleration process during HILDCAAs. Pitch angle distribution showing stronger enhancement for nearly $90^\circ$ particles and energization of these particles reaching higher energies for more extended HILDCAA events.

During both events, whistler-mode chorus waves show enhanced power, previously identified as key contributors to local electron acceleration, which may act in synergy with ULF waves, accelerating low-energy electrons \citep{baker2014relativistic, kletzing2013electric}. Chorus emissions arise from cyclotron resonance instability in the low-density region beyond the plasmapause. They are triggered by the injection of medium-energy ($\sim 1–30$ keV) electrons into the inner magnetosphere during substorms or periods of increased convection \citep{thorne2021wave}. The MagISE shows multiple injections of such low-energy electrons during the studied HILDCAAs, indicating a conducive environment for chorus wave generation.  Moreover, the strong correlation between elevated Ultra-Low Frequency (ULF) wave activity and enhanced electron acceleration during HILDCAA events points to the role of wave-particle interaction. ULF waves facilitate the radial transport and energization of radiation belt electrons through diffusion and resonant interactions, particularly during prolonged geomagnetic activity \citep{tsurutani1997some,tsurutani2004high,hajra2024ultra}. The sustained presence of ULF waves in long-duration events suggests that these processes possibly drive continuous electron acceleration, leading to persistent high-energy populations in the outer belt.

In summary, this study highlights the crucial role of the duration of the HILDCAA event in shaping the dynamics of the Earth's outer radiation belt. Our analysis of two contrasting events, a short- and a long-duration event, reveals that while both enhance electron energy and flux, their temporal evolution and intensity differ significantly. The short event triggered rapid but transient increases, whereas the long event led to sustained and more pronounced enhancements. Additionally, the strong correlation between elevated ULF and chorus wave activity and increased electron acceleration underscores the importance of wave-particle interactions, especially during prolonged events. These findings emphasize the need to consider the duration of events in the prediction and assessment of space weather. Future research should further investigate the underlying physical processes with multi-point satellite observations and modeling to improve forecasting and mitigate the effects of space weather on space-based technological systems.

%
%

\section*{Open Research Section}
The data for the present study can be obtained from:
\begin{enumerate}
    \item \url{https://omniweb.gsfc.nasa.gov/html/about_data.html}
    \item For REPT, EMFISIS and MagEIS data, \url{https://rbspgway.jhuapl.edu/} 
\end{enumerate}

\acknowledgments
The authors acknowledge the use of data provided by NASA OmniWeb, EMFISIS, MagEIS, and REPT instruments onboard Van Allen Probes, WDC for Geomagnetism, Kyoto, Japan, and IMAGE. A.N. thanks MHRD for financial support and the Sardar Vallabhbhai National Institute of Technology for research facilities. A.N. also thanks ISRO/Vikram Sarabhai Space Centre, India, for providing a Ph.D. internship opportunity to facilitate the current research.

%
%

\bibliography{agu}

\begin{thebibliography}{}

\bibitem [\protect \citeauthoryear {%
Akasofu%
}{%
Akasofu%
}{%
{\protect \APACyear {1981}}%
}]{%
akasofu1981relationships}
\APACinsertmetastar {%
akasofu1981relationships}%
\begin{APACrefauthors}%
Akasofu, S\BHBI I.%
\end{APACrefauthors}%
\unskip\
\newblock
\APACrefYearMonthDay{1981}{}{}.
\newblock
{\BBOQ}\APACrefatitle {Relationships between the AE and Dst indices during geomagnetic storms} {Relationships between the ae and dst indices during geomagnetic storms}.{\BBCQ}
\newblock
\APACjournalVolNumPages{Journal of Geophysical Research: Space Physics}{86}{A6}{4820--4822}.
\PrintBackRefs{\CurrentBib}

\bibitem [\protect \citeauthoryear {%
Baker%
, Kanekal%
\BCBL {}\ \protect \BOthers {.}}{%
Baker%
\ \protect \BOthers {.}}{%
{\protect \APACyear {2014}}%
}]{%
baker2014relativistic}
\APACinsertmetastar {%
baker2014relativistic}%
\begin{APACrefauthors}%
Baker, D\BPBI N.%
, Kanekal, S.%
\BCBL {}\ \BOthersPeriod {.}\end{APACrefauthors}%
\unskip\
\newblock
\APACrefYearMonthDay{2014}{}{}.
\newblock
{\BBOQ}\APACrefatitle {The Relativistic Electron-Proton Telescope (REPT) instrument on board the Radiation Belt Storm Probes (RBSP) spacecraft: Characterization of Earth’s radiation belt high-energy particle populations} {The relativistic electron-proton telescope (rept) instrument on board the radiation belt storm probes (rbsp) spacecraft: Characterization of earth’s radiation belt high-energy particle populations}.{\BBCQ}
\newblock
\APACjournalVolNumPages{The van allen probes mission}{}{}{337--381}.
\PrintBackRefs{\CurrentBib}

\bibitem [\protect \citeauthoryear {%
Baker%
, Kanekal%
\BCBL {}\ \protect \BOthers {.}}{%
Baker%
\ \protect \BOthers {.}}{%
{\protect \APACyear {2021}}%
}]{%
baker2021relativistic}
\APACinsertmetastar {%
baker2021relativistic}%
\begin{APACrefauthors}%
Baker, D\BPBI N.%
, Kanekal, S\BPBI G.%
\BCBL {}\ \BOthersPeriod {.}\end{APACrefauthors}%
\unskip\
\newblock
\APACrefYearMonthDay{2021}{}{}.
\newblock
{\BBOQ}\APACrefatitle {The relativistic electron-proton telescope (REPT) investigation: Design, operational properties, and science highlights} {The relativistic electron-proton telescope (rept) investigation: Design, operational properties, and science highlights}.{\BBCQ}
\newblock
\APACjournalVolNumPages{Space science reviews}{217}{5}{68}.
\PrintBackRefs{\CurrentBib}

\bibitem [\protect \citeauthoryear {%
Blake%
, Carranza%
\BCBL {}\ \protect \BOthers {.}}{%
Blake%
\ \protect \BOthers {.}}{%
{\protect \APACyear {2014}}%
}]{%
blake2014magnetic}
\APACinsertmetastar {%
blake2014magnetic}%
\begin{APACrefauthors}%
Blake, J.%
, Carranza, P.%
\BCBL {}\ \BOthersPeriod {.}\end{APACrefauthors}%
\unskip\
\newblock
\APACrefYearMonthDay{2014}{}{}.
\newblock
{\BBOQ}\APACrefatitle {The magnetic electron ion spectrometer (MagEIS) instruments aboard the radiation belt storm probes (RBSP) spacecraft} {The magnetic electron ion spectrometer (mageis) instruments aboard the radiation belt storm probes (rbsp) spacecraft}.{\BBCQ}
\newblock
\APACjournalVolNumPages{The van allen probes mission}{}{}{383--421}.
\PrintBackRefs{\CurrentBib}

\bibitem [\protect \citeauthoryear {%
Claudepierre%
, O'Brien%
\BCBL {}\ \protect \BOthers {.}}{%
Claudepierre%
\ \protect \BOthers {.}}{%
{\protect \APACyear {2015}}%
}]{%
claudepierre2015background}
\APACinsertmetastar {%
claudepierre2015background}%
\begin{APACrefauthors}%
Claudepierre, S.%
, O'Brien, T.%
\BCBL {}\ \BOthersPeriod {.}\end{APACrefauthors}%
\unskip\
\newblock
\APACrefYearMonthDay{2015}{}{}.
\newblock
{\BBOQ}\APACrefatitle {A background correction algorithm for Van Allen Probes MagEIS electron flux measurements} {A background correction algorithm for van allen probes mageis electron flux measurements}.{\BBCQ}
\newblock
\APACjournalVolNumPages{Journal of Geophysical Research: Space Physics}{120}{7}{5703--5727}.
\PrintBackRefs{\CurrentBib}

\bibitem [\protect \citeauthoryear {%
Da~Silva%
, Sibeck%
\BCBL {}\ \protect \BOthers {.}}{%
Da~Silva%
\ \protect \BOthers {.}}{%
{\protect \APACyear {2019}}%
}]{%
da2019contribution}
\APACinsertmetastar {%
da2019contribution}%
\begin{APACrefauthors}%
Da~Silva, L\BPBI A.%
, Sibeck, D.%
\BCBL {}\ \BOthersPeriod {.}\end{APACrefauthors}%
\unskip\
\newblock
\APACrefYearMonthDay{2019}{}{}.
\newblock
{\BBOQ}\APACrefatitle {Contribution of ULF wave activity to the global recovery of the outer radiation belt during the passage of a high-speed solar wind stream observed in September 2014} {Contribution of ulf wave activity to the global recovery of the outer radiation belt during the passage of a high-speed solar wind stream observed in september 2014}.{\BBCQ}
\newblock
\APACjournalVolNumPages{Journal of Geophysical Research: Space Physics}{124}{3}{1660--1678}.
\PrintBackRefs{\CurrentBib}

\bibitem [\protect \citeauthoryear {%
Davis%
\ \BBA {} Sugiura%
}{%
Davis%
\ \BBA {} Sugiura%
}{%
{\protect \APACyear {1966}}%
}]{%
davis1966auroral}
\APACinsertmetastar {%
davis1966auroral}%
\begin{APACrefauthors}%
Davis, T\BPBI N.%
\BCBT {}\ \BBA {} Sugiura, M.%
\end{APACrefauthors}%
\unskip\
\newblock
\APACrefYearMonthDay{1966}{}{}.
\newblock
{\BBOQ}\APACrefatitle {Auroral electrojet activity index AE and its universal time variations} {Auroral electrojet activity index ae and its universal time variations}.{\BBCQ}
\newblock
\APACjournalVolNumPages{Journal of Geophysical Research}{71}{3}{785--801}.
\PrintBackRefs{\CurrentBib}

\bibitem [\protect \citeauthoryear {%
Fennell%
, Claudepierre%
\BCBL {}\ \protect \BOthers {.}}{%
Fennell%
\ \protect \BOthers {.}}{%
{\protect \APACyear {2015}}%
}]{%
fennell2015van}
\APACinsertmetastar {%
fennell2015van}%
\begin{APACrefauthors}%
Fennell, J.%
, Claudepierre, S.%
\BCBL {}\ \BOthersPeriod {.}\end{APACrefauthors}%
\unskip\
\newblock
\APACrefYearMonthDay{2015}{}{}.
\newblock
{\BBOQ}\APACrefatitle {Van Allen Probes show that the inner radiation zone contains no MeV electrons: ECT/MagEIS data} {Van allen probes show that the inner radiation zone contains no mev electrons: Ect/mageis data}.{\BBCQ}
\newblock
\APACjournalVolNumPages{Geophysical Research Letters}{42}{5}{1283--1289}.
\PrintBackRefs{\CurrentBib}

\bibitem [\protect \citeauthoryear {%
Gonzalez%
, Joselyn%
\BCBL {}\ \protect \BOthers {.}}{%
Gonzalez%
\ \protect \BOthers {.}}{%
{\protect \APACyear {1994}}%
}]{%
gonzalez1994geomagnetic}
\APACinsertmetastar {%
gonzalez1994geomagnetic}%
\begin{APACrefauthors}%
Gonzalez, W.%
, Joselyn, J\BHBI A.%
\BCBL {}\ \BOthersPeriod {.}\end{APACrefauthors}%
\unskip\
\newblock
\APACrefYearMonthDay{1994}{}{}.
\newblock
{\BBOQ}\APACrefatitle {What is a geomagnetic storm?} {What is a geomagnetic storm?}{\BBCQ}
\newblock
\APACjournalVolNumPages{Journal of Geophysical Research: Space Physics}{99}{A4}{5771--5792}.
\PrintBackRefs{\CurrentBib}

\bibitem [\protect \citeauthoryear {%
Hajra%
, Echer%
\BCBL {}\ \protect \BOthers {.}}{%
Hajra%
\ \protect \BOthers {.}}{%
{\protect \APACyear {2013}}%
}]{%
hajra2013solar}
\APACinsertmetastar {%
hajra2013solar}%
\begin{APACrefauthors}%
Hajra, R.%
, Echer, E.%
\BCBL {}\ \BOthersPeriod {.}\end{APACrefauthors}%
\unskip\
\newblock
\APACrefYearMonthDay{2013}{}{}.
\newblock
{\BBOQ}\APACrefatitle {Solar cycle dependence of High-Intensity Long-Duration Continuous AE Activity (HILDCAA) events, relativistic electron predictors?} {Solar cycle dependence of high-intensity long-duration continuous ae activity (hildcaa) events, relativistic electron predictors?}{\BBCQ}
\newblock
\APACjournalVolNumPages{Journal of Geophysical Research: Space Physics}{118}{9}{5626--5638}.
\PrintBackRefs{\CurrentBib}

\bibitem [\protect \citeauthoryear {%
Hajra%
, Tsurutani%
\BCBL {}\ \protect \BOthers {.}}{%
Hajra%
\ \protect \BOthers {.}}{%
{\protect \APACyear {2014}}%
}]{%
hajra2014relativistic}
\APACinsertmetastar {%
hajra2014relativistic}%
\begin{APACrefauthors}%
Hajra, R.%
, Tsurutani, B\BPBI T.%
\BCBL {}\ \BOthersPeriod {.}\end{APACrefauthors}%
\unskip\
\newblock
\APACrefYearMonthDay{2014}{}{}.
\newblock
{\BBOQ}\APACrefatitle {Relativistic electron acceleration during high-intensity, long-duration, continuous AE activity (HILDCAA) events: Solar cycle phase dependences} {Relativistic electron acceleration during high-intensity, long-duration, continuous ae activity (hildcaa) events: Solar cycle phase dependences}.{\BBCQ}
\newblock
\APACjournalVolNumPages{Geophysical Research Letters}{41}{6}{1876--1881}.
\PrintBackRefs{\CurrentBib}

\bibitem [\protect \citeauthoryear {%
Hajra%
, Tsurutani%
\BCBL {}\ \protect \BOthers {.}}{%
Hajra%
\ \protect \BOthers {.}}{%
{\protect \APACyear {2015}}%
}]{%
hajra2015relativistic}
\APACinsertmetastar {%
hajra2015relativistic}%
\begin{APACrefauthors}%
Hajra, R.%
, Tsurutani, B\BPBI T.%
\BCBL {}\ \BOthersPeriod {.}\end{APACrefauthors}%
\unskip\
\newblock
\APACrefYearMonthDay{2015}{}{}.
\newblock
{\BBOQ}\APACrefatitle {Relativistic (E> 0.6,> 2.0, and> 4.0 MeV) electron acceleration at geosynchronous orbit during high-intensity, long-duration, continuous AE activity (HILDCAA) events} {Relativistic (e> 0.6,> 2.0, and> 4.0 mev) electron acceleration at geosynchronous orbit during high-intensity, long-duration, continuous ae activity (hildcaa) events}.{\BBCQ}
\newblock
\APACjournalVolNumPages{The Astrophysical Journal}{799}{1}{39}.
\PrintBackRefs{\CurrentBib}

\bibitem [\protect \citeauthoryear {%
Hajra%
, Tsurutani%
\BCBL {}\ \protect \BOthers {.}}{%
Hajra%
\ \protect \BOthers {.}}{%
{\protect \APACyear {2024}}%
}]{%
hajra2024ultra}
\APACinsertmetastar {%
hajra2024ultra}%
\begin{APACrefauthors}%
Hajra, R.%
, Tsurutani, B\BPBI T.%
\BCBL {}\ \BOthersPeriod {.}\end{APACrefauthors}%
\unskip\
\newblock
\APACrefYearMonthDay{2024}{}{}.
\newblock
{\BBOQ}\APACrefatitle {Ultra-relativistic Electron Acceleration during High-intensity Long-duration Continuous Auroral Electrojet Activity Events} {Ultra-relativistic electron acceleration during high-intensity long-duration continuous auroral electrojet activity events}.{\BBCQ}
\newblock
\APACjournalVolNumPages{The Astrophysical Journal}{965}{2}{146}.
\PrintBackRefs{\CurrentBib}

\bibitem [\protect \citeauthoryear {%
Hapgood%
}{%
Hapgood%
}{%
{\protect \APACyear {1992}}%
}]{%
hapgood1992space}
\APACinsertmetastar {%
hapgood1992space}%
\begin{APACrefauthors}%
Hapgood, M.%
\end{APACrefauthors}%
\unskip\
\newblock
\APACrefYearMonthDay{1992}{}{}.
\newblock
{\BBOQ}\APACrefatitle {Space physics coordinate transformations: A user guide} {Space physics coordinate transformations: A user guide}.{\BBCQ}
\newblock
\APACjournalVolNumPages{Planetary and Space Science}{40}{5}{711--717}.
\PrintBackRefs{\CurrentBib}

\bibitem [\protect \citeauthoryear {%
Iles%
, Meredith%
, Fazakerley%
\BCBL {}\ \BBA {} Horne%
}{%
Iles%
\ \protect \BOthers {.}}{%
{\protect \APACyear {2006}}%
}]{%
iles2006phase}
\APACinsertmetastar {%
iles2006phase}%
\begin{APACrefauthors}%
Iles, R\BPBI H.%
, Meredith, N\BPBI P.%
, Fazakerley, A\BPBI N.%
\BCBL {}\ \BBA {} Horne, R\BPBI B.%
\end{APACrefauthors}%
\unskip\
\newblock
\APACrefYearMonthDay{2006}{}{}.
\newblock
{\BBOQ}\APACrefatitle {Phase space density analysis of the outer radiation belt energetic electron dynamics} {Phase space density analysis of the outer radiation belt energetic electron dynamics}.{\BBCQ}
\newblock
\APACjournalVolNumPages{Journal of Geophysical Research: Space Physics}{111}{A3}{}.
\PrintBackRefs{\CurrentBib}

\bibitem [\protect \citeauthoryear {%
Kletzing%
, Kurth%
\BCBL {}\ \protect \BOthers {.}}{%
Kletzing%
\ \protect \BOthers {.}}{%
{\protect \APACyear {2013}}%
}]{%
kletzing2013electric}
\APACinsertmetastar {%
kletzing2013electric}%
\begin{APACrefauthors}%
Kletzing, C.%
, Kurth, W.%
\BCBL {}\ \BOthersPeriod {.}\end{APACrefauthors}%
\unskip\
\newblock
\APACrefYearMonthDay{2013}{}{}.
\newblock
{\BBOQ}\APACrefatitle {The electric and magnetic field instrument suite and integrated science (EMFISIS) on RBSP} {The electric and magnetic field instrument suite and integrated science (emfisis) on rbsp}.{\BBCQ}
\newblock
\APACjournalVolNumPages{Space Science Reviews}{179}{}{127--181}.
\PrintBackRefs{\CurrentBib}

\bibitem [\protect \citeauthoryear {%
Laundal%
\ \BBA {} Richmond%
}{%
Laundal%
\ \BBA {} Richmond%
}{%
{\protect \APACyear {2017}}%
}]{%
laundal2017magnetic}
\APACinsertmetastar {%
laundal2017magnetic}%
\begin{APACrefauthors}%
Laundal, K\BPBI M.%
\BCBT {}\ \BBA {} Richmond, A\BPBI D.%
\end{APACrefauthors}%
\unskip\
\newblock
\APACrefYearMonthDay{2017}{}{}.
\newblock
{\BBOQ}\APACrefatitle {Magnetic coordinate systems} {Magnetic coordinate systems}.{\BBCQ}
\newblock
\APACjournalVolNumPages{Space Science Reviews}{206}{1}{27--59}.
\PrintBackRefs{\CurrentBib}

\bibitem [\protect \citeauthoryear {%
Mauk%
, Fox%
\BCBL {}\ \protect \BOthers {.}}{%
Mauk%
\ \protect \BOthers {.}}{%
{\protect \APACyear {2014}}%
}]{%
mauk2014science}
\APACinsertmetastar {%
mauk2014science}%
\begin{APACrefauthors}%
Mauk, B.%
, Fox, N\BPBI J.%
\BCBL {}\ \BOthersPeriod {.}\end{APACrefauthors}%
\unskip\
\newblock
\APACrefYearMonthDay{2014}{}{}.
\newblock
{\BBOQ}\APACrefatitle {Science objectives and rationale for the radiation belt storm probes mission} {Science objectives and rationale for the radiation belt storm probes mission}.{\BBCQ}
\newblock
\APACjournalVolNumPages{The van Allen probes mission}{}{}{3--27}.
\PrintBackRefs{\CurrentBib}

\bibitem [\protect \citeauthoryear {%
Morley%
, Henderson%
\BCBL {}\ \protect \BOthers {.}}{%
Morley%
\ \protect \BOthers {.}}{%
{\protect \APACyear {2013}}%
}]{%
morley2013phase}
\APACinsertmetastar {%
morley2013phase}%
\begin{APACrefauthors}%
Morley, S.%
, Henderson, M.%
\BCBL {}\ \BOthersPeriod {.}\end{APACrefauthors}%
\unskip\
\newblock
\APACrefYearMonthDay{2013}{}{}.
\newblock
{\BBOQ}\APACrefatitle {Phase space density matching of relativistic electrons using the Van Allen Probes: REPT results} {Phase space density matching of relativistic electrons using the van allen probes: Rept results}.{\BBCQ}
\newblock
\APACjournalVolNumPages{Geophysical Research Letters}{40}{18}{4798--4802}.
\PrintBackRefs{\CurrentBib}

\bibitem [\protect \citeauthoryear {%
Nema%
, Bhaskar%
\BCBL {}\ \protect \BOthers {.}}{%
Nema%
\ \protect \BOthers {.}}{%
{\protect \APACyear {2024}}%
}]{%
nema2024impact}
\APACinsertmetastar {%
nema2024impact}%
\begin{APACrefauthors}%
Nema, A.%
, Bhaskar, A.%
\BCBL {}\ \BOthersPeriod {.}\end{APACrefauthors}%
\unskip\
\newblock
\APACrefYearMonthDay{2024}{}{}.
\newblock
{\BBOQ}\APACrefatitle {Impact of High Intensity Long-Duration Continuous Auroral Electrojet Activity (HILDCAAs) on relativistic electrons of the radiation belt of Earth during Van Allen probe era} {Impact of high intensity long-duration continuous auroral electrojet activity (hildcaas) on relativistic electrons of the radiation belt of earth during van allen probe era}.{\BBCQ}
\newblock
\APACjournalVolNumPages{arXiv preprint arXiv:2410.20442}{}{}{}.
\PrintBackRefs{\CurrentBib}

\bibitem [\protect \citeauthoryear {%
Roederer%
}{%
Roederer%
}{%
{\protect \APACyear {2012}}%
}]{%
roederer2012dynamics}
\APACinsertmetastar {%
roederer2012dynamics}%
\begin{APACrefauthors}%
Roederer, J\BPBI G.%
\end{APACrefauthors}%
\unskip\
\newblock
\APACrefYear{2012}.
\newblock
\APACrefbtitle {Dynamics of geomagnetically trapped radiation} {Dynamics of geomagnetically trapped radiation}\ (\BVOL~2).
\newblock
\APACaddressPublisher{}{Springer Science \& Business Media}.
\PrintBackRefs{\CurrentBib}

\bibitem [\protect \citeauthoryear {%
Rostoker%
}{%
Rostoker%
}{%
{\protect \APACyear {1972}}%
}]{%
rostoker1972geomagnetic}
\APACinsertmetastar {%
rostoker1972geomagnetic}%
\begin{APACrefauthors}%
Rostoker, G.%
\end{APACrefauthors}%
\unskip\
\newblock
\APACrefYearMonthDay{1972}{}{}.
\newblock
{\BBOQ}\APACrefatitle {Geomagnetic indices} {Geomagnetic indices}.{\BBCQ}
\newblock
\APACjournalVolNumPages{Reviews of Geophysics}{10}{4}{935--950}.
\PrintBackRefs{\CurrentBib}

\bibitem [\protect \citeauthoryear {%
Runov%
, Angelopoulos%
\BCBL {}\ \protect \BOthers {.}}{%
Runov%
\ \protect \BOthers {.}}{%
{\protect \APACyear {2025}}%
}]{%
runov2025themis}
\APACinsertmetastar {%
runov2025themis}%
\begin{APACrefauthors}%
Runov, A.%
, Angelopoulos, V.%
\BCBL {}\ \BOthersPeriod {.}\end{APACrefauthors}%
\unskip\
\newblock
\APACrefYearMonthDay{2025}{}{}.
\newblock
{\BBOQ}\APACrefatitle {THEMIS observations of relativistic electrons at the nightside transition region during HILDCAA events} {Themis observations of relativistic electrons at the nightside transition region during hildcaa events}.{\BBCQ}
\newblock
\APACjournalVolNumPages{Journal of Geophysical Research: Space Physics}{130}{2}{e2024JA033179}.
\PrintBackRefs{\CurrentBib}

\bibitem [\protect \citeauthoryear {%
Russell%
}{%
Russell%
}{%
{\protect \APACyear {1971}}%
}]{%
russell1971geophysical}
\APACinsertmetastar {%
russell1971geophysical}%
\begin{APACrefauthors}%
Russell, C\BPBI T.%
\end{APACrefauthors}%
\unskip\
\newblock
\APACrefYearMonthDay{1971}{}{}.
\newblock
{\BBOQ}\APACrefatitle {Geophysical coordinate transformations} {Geophysical coordinate transformations}.{\BBCQ}
\newblock
\APACjournalVolNumPages{Cosmic electrodynamics}{2}{2}{184--196}.
\PrintBackRefs{\CurrentBib}

\bibitem [\protect \citeauthoryear {%
Russell%
, McPherron%
\BCBL {}\ \protect \BOthers {.}}{%
Russell%
\ \protect \BOthers {.}}{%
{\protect \APACyear {1974}}%
}]{%
russell1974cause}
\APACinsertmetastar {%
russell1974cause}%
\begin{APACrefauthors}%
Russell, C\BPBI T.%
, McPherron, R\BPBI L.%
\BCBL {}\ \BOthersPeriod {.}\end{APACrefauthors}%
\unskip\
\newblock
\APACrefYearMonthDay{1974}{}{}.
\newblock
{\BBOQ}\APACrefatitle {On the cause of geomagnetic storms} {On the cause of geomagnetic storms}.{\BBCQ}
\newblock
\APACjournalVolNumPages{Journal of Geophysical Research}{79}{7}{1105--1109}.
\PrintBackRefs{\CurrentBib}

\bibitem [\protect \citeauthoryear {%
Schiller%
, Tu%
\BCBL {}\ \protect \BOthers {.}}{%
Schiller%
\ \protect \BOthers {.}}{%
{\protect \APACyear {2017}}%
}]{%
schiller2017simultaneous}
\APACinsertmetastar {%
schiller2017simultaneous}%
\begin{APACrefauthors}%
Schiller, Q.%
, Tu, W.%
\BCBL {}\ \BOthersPeriod {.}\end{APACrefauthors}%
\unskip\
\newblock
\APACrefYearMonthDay{2017}{}{}.
\newblock
{\BBOQ}\APACrefatitle {Simultaneous event-specific estimates of transport, loss, and source rates for relativistic outer radiation belt electrons} {Simultaneous event-specific estimates of transport, loss, and source rates for relativistic outer radiation belt electrons}.{\BBCQ}
\newblock
\APACjournalVolNumPages{Journal of Geophysical Research: Space Physics}{122}{3}{3354--3373}.
\PrintBackRefs{\CurrentBib}

\bibitem [\protect \citeauthoryear {%
Schulz%
\ \BBA {} Lanzerotti%
}{%
Schulz%
\ \BBA {} Lanzerotti%
}{%
{\protect \APACyear {2012}}%
}]{%
schulz2012particle}
\APACinsertmetastar {%
schulz2012particle}%
\begin{APACrefauthors}%
Schulz, M.%
\BCBT {}\ \BBA {} Lanzerotti, L\BPBI J.%
\end{APACrefauthors}%
\unskip\
\newblock
\APACrefYear{2012}.
\newblock
\APACrefbtitle {Particle diffusion in the radiation belts} {Particle diffusion in the radiation belts}\ (\BVOL~7).
\newblock
\APACaddressPublisher{}{Springer Science \& Business Media}.
\PrintBackRefs{\CurrentBib}

\bibitem [\protect \citeauthoryear {%
Thorne%
, Bortnik%
, Li%
\BCBL {}\ \BBA {} Ma%
}{%
Thorne%
\ \protect \BOthers {.}}{%
{\protect \APACyear {2021}}%
}]{%
thorne2021wave}
\APACinsertmetastar {%
thorne2021wave}%
\begin{APACrefauthors}%
Thorne, R\BPBI M.%
, Bortnik, J.%
, Li, W.%
\BCBL {}\ \BBA {} Ma, Q.%
\end{APACrefauthors}%
\unskip\
\newblock
\APACrefYearMonthDay{2021}{}{}.
\newblock
{\BBOQ}\APACrefatitle {Wave--particle interactions in the Earth's magnetosphere} {Wave--particle interactions in the earth's magnetosphere}.{\BBCQ}
\newblock
\APACjournalVolNumPages{Magnetospheres in the solar system}{}{}{93--108}.
\PrintBackRefs{\CurrentBib}

\bibitem [\protect \citeauthoryear {%
Tsurutani%
\ \BBA {} Gonzalez%
}{%
Tsurutani%
\ \BBA {} Gonzalez%
}{%
{\protect \APACyear {1987}}%
}]{%
tsurutani1987cause}
\APACinsertmetastar {%
tsurutani1987cause}%
\begin{APACrefauthors}%
Tsurutani, B\BPBI T.%
\BCBT {}\ \BBA {} Gonzalez, W\BPBI D.%
\end{APACrefauthors}%
\unskip\
\newblock
\APACrefYearMonthDay{1987}{}{}.
\newblock
{\BBOQ}\APACrefatitle {The cause of high-intensity long-duration continuous AE activity (HILDCAAs): Interplanetary Alfv{\'e}n wave trains} {The cause of high-intensity long-duration continuous ae activity (hildcaas): Interplanetary alfv{\'e}n wave trains}.{\BBCQ}
\newblock
\APACjournalVolNumPages{Planetary and Space Science}{35}{4}{405--412}.
\PrintBackRefs{\CurrentBib}

\bibitem [\protect \citeauthoryear {%
Tsurutani%
, Gonzalez%
\BCBL {}\ \protect \BOthers {.}}{%
Tsurutani%
\ \protect \BOthers {.}}{%
{\protect \APACyear {2004}}%
}]{%
tsurutani2004high}
\APACinsertmetastar {%
tsurutani2004high}%
\begin{APACrefauthors}%
Tsurutani, B\BPBI T.%
, Gonzalez, W\BPBI D.%
\BCBL {}\ \BOthersPeriod {.}\end{APACrefauthors}%
\unskip\
\newblock
\APACrefYearMonthDay{2004}{}{}.
\newblock
{\BBOQ}\APACrefatitle {Are high-intensity long-duration continuous AE activity (HILDCAA) events substorm expansion events?} {Are high-intensity long-duration continuous ae activity (hildcaa) events substorm expansion events?}{\BBCQ}
\newblock
\APACjournalVolNumPages{Journal of Atmospheric and Solar-Terrestrial Physics}{66}{2}{167--176}.
\PrintBackRefs{\CurrentBib}

\bibitem [\protect \citeauthoryear {%
Tsurutani%
\ \BBA {} Lakhina%
}{%
Tsurutani%
\ \BBA {} Lakhina%
}{%
{\protect \APACyear {1997}}%
}]{%
tsurutani1997some}
\APACinsertmetastar {%
tsurutani1997some}%
\begin{APACrefauthors}%
Tsurutani, B\BPBI T.%
\BCBT {}\ \BBA {} Lakhina, G\BPBI S.%
\end{APACrefauthors}%
\unskip\
\newblock
\APACrefYearMonthDay{1997}{}{}.
\newblock
{\BBOQ}\APACrefatitle {Some basic concepts of wave-particle interactions in collisionless plasmas} {Some basic concepts of wave-particle interactions in collisionless plasmas}.{\BBCQ}
\newblock
\APACjournalVolNumPages{Reviews of Geophysics}{35}{4}{491--501}.
\PrintBackRefs{\CurrentBib}

\bibitem [\protect \citeauthoryear {%
Tsyganenko%
\ \BBA {} Sitnov%
}{%
Tsyganenko%
\ \BBA {} Sitnov%
}{%
{\protect \APACyear {2005}}%
}]{%
tsyganenko2005modeling}
\APACinsertmetastar {%
tsyganenko2005modeling}%
\begin{APACrefauthors}%
Tsyganenko, N.%
\BCBT {}\ \BBA {} Sitnov, M.%
\end{APACrefauthors}%
\unskip\
\newblock
\APACrefYearMonthDay{2005}{}{}.
\newblock
{\BBOQ}\APACrefatitle {Modeling the dynamics of the inner magnetosphere during strong geomagnetic storms} {Modeling the dynamics of the inner magnetosphere during strong geomagnetic storms}.{\BBCQ}
\newblock
\APACjournalVolNumPages{Journal of Geophysical Research: Space Physics}{110}{A3}{}.
\PrintBackRefs{\CurrentBib}

\end{thebibliography}

\end{document}